\documentclass[11pt]{article}
\usepackage[english]{babel}
\usepackage{pdflscape}
\usepackage{fancyhdr} 
\usepackage{amsthm,amsmath,amssymb,eucal}
\usepackage{graphicx, color}
\usepackage{caption}
\usepackage{float}
\usepackage[T1]{fontenc}
\usepackage[numbered,framed]{matlab-prettifier}
\usepackage{pgffor}
\usepackage{mathtools}
\usepackage{pdfpages}
\theoremstyle{definition}

\theoremstyle{plain}

\usepackage{pdfpages}
\usepackage[style=nature]{biblatex}
\usepackage[letterpaper,top=2cm,bottom=2cm,left=3cm,right=3cm,marginparwidth=1.75cm]{geometry}

\usepackage{amsmath}
\usepackage{graphicx}
\usepackage[colorlinks=true, allcolors=blue]{hyperref}
\title{Supporting Information: Emergent kinetics of in vitro transcription from interactions of T7 RNA polymerase and DNA}
\author{Nathan M. Stover†, Marieke de Bock†, Julie Chen, Jacob Rosenfeld,\\ Maria del Carme Pons Royo, Allan S. Myerson, and Richard D. Braatz* \\ \\ stover@mit.edu, braatz@mit.edu \\ 
†These authors contributed equally to this work\\
* Corresponding author}
\date{}
\begin{document}


# Emergent kinetics of in vitro transcription from interactions of T7 RNA polymerase and DNA

Nathan M. Stover[†], Marieke De Bock[†], Julie Chen, Jacob Rosenfeld,
Maria del Carme Pons Royo, Allan S. Myerson, and Richard D. Braatz[*]

*Department of Chemical Engineering, Massachusetts Institute of Technology, Cambridge, MA*

[†]These authors contributed equally to this work

[*] Corresponding author

E-mail: stover@mit.edu, braatz@mit.edu

## Abstract

The in vitro transcription reaction (IVT) is of growing importance for the manufacture of RNA vaccines and therapeutics. While the kinetics of the microscopic steps of this reaction (promoter binding, initiation, and elongation) are well studied, the rate law of overall RNA synthesis that emerges from this system is unclear. In this work, we show that a model that incorporates both initiation and elongation steps is essential for describing trends in IVT kinetics in conditions relevant to RNA manufacturing. In contrast to previous reports, we find that the IVT reaction can be either initiation- or elongation-limited depending on solution conditions. This initiation-elongation model is also essential for describing the effect of salts, which disrupt polymerase-promoter binding, on transcription rates. Polymerase-polymerase interactions during elongation are incorporated into our modeling framework and found to have nonzero but unidentifiable effects on macroscopic transcription rates. Finally, we develop an extension of our modeling approach to quantitatively describe and experimentally evaluate RNA- and DNA-templated mechanisms for the formation of double-stranded RNA (dsRNA) impurities. We show experimental results that indicate that an RNA-templated mechanism is not appropriate for describing macroscopic dsRNA formation in the context of RNA manufacturing.

**Introduction**

The in vitro transcription (IVT) reaction for synthesis of RNA is a necessary step for the production of a growing number of vaccines and therapeutics. IVT is a cell-free biochemical reaction that requires a DNA template and a DNA-dependent RNA polymerase enzyme, most commonly T7 RNA polymerase. A kinetic law for the rate of RNA synthesis as a function of the concentrations of these catalysts can aid in design, optimization, and mechanistic understanding of the IVT process. However, a rate law for IVT in conditions relevant to RNA manufacturing has not been fully developed.

The elementary kinetic mechanisms that constitute IVT are well studied. Each of the promoter binding, initiation, and elongation steps has been studied using a diverse and orthogonal set of tools, including thermodynamic measurements[1,2], kinetic modeling[3–6], structural analysis[7,8], and single-molecule experiments[9–11]. Despite this microscopic understanding, there is little research into the emergent macroscopic kinetics of systems in which these steps coexist. This knowledge gap is relevant in the context of manufacturing long RNA sequences (>1000 bp), such as mRNA and self-amplifying RNA (saRNA) vaccines.

The work of Arnold et al.[12] is the most complete past approach to modeling macroscopic transcription rates. Arnold et al. report that the estimated elongation rate constant is much greater than the estimated initiation rate constant and conclude that the IVT reaction is initiation limited for all industrially relevant sequence lengths. However, this work is incomplete and not representative of the IVT reaction in a biomanufacturing context. As will be described in the results section, the kinetic model used by Arnold et al.[1] is not appropriate for the synthesis of long RNA sequences. Secondly, they estimated transcription and elongation rate constants using a poorly defined parameter estimation approach in which the number of fitted parameters nearly equaled the number of data points. It is well understood that this approach can be extremely sensitive to experimental noise or out-of-model effects. As a consequence, their estimated elongation rate constant ($2\times10^{11}$ 1/h) is not only far greater than orthogonal estimates from single-molecule studies (~$5\times10^5$ 1/h)[10], but is $10^3$ times greater than the diffusion-limited rate constant, which implies that it is an aphysical artifact of errors in the parameter estimation process. These considerations are relevant for practitioners in the field of RNA manufacturing. For example, Boman et al.[14] rely on Arnold et al.'s conclusion of initiation limitation to estimate the effect of sequence length on IVT rates for process development. A kinetic

---

[1] First developed by Pozhitkov et al.[13]

modeling approach that appropriately incorporates all steps of the transcription process can serve as a useful tool for several aspects of RNA manufacturing, including accelerating process development of new sequences and designing dynamic models of IVT.

Beyond predicting the rates of product RNA formation, kinetic modeling can be a useful tool for understanding the formation of double-stranded RNA (dsRNA) impurities. Double-stranded RNA is a highly immunogenic byproduct of IVT which is costly to remove in downstream purification. Multiple mechanisms have been proposed to describe the formation of dsRNA, including RNA-templated 3' addition[15] and DNA-templated synthesis of antisense RNA followed by hybridization[16]. While both of these mechanisms have been experimentally observed within the context of model systems, their usefulness for the quantitative design of the IVT reaction is unclear and there are no publications on attempting to quantify the kinetic predictions of these mechanisms.

In this work, we investigate a kinetic rate law that incorporates polymerase-promoter binding, initiation, and elongation steps from a first-principles standpoint. We show that this initiation-elongation model is necessary for describing the rate of transcription in regimes relevant to the manufacturing of mRNA. We demonstrate how the kinetic parameters of this model can be estimated from a minimal set of experiments, which allows for a comparative analysis between different DNA sequences and serves as a guide for practitioners on troubleshooting and understanding the application of IVT to novel sequences. We consider both the effect of polymerase-polymerase interactions and polymerase-DNA binding disruptions on the kinetic predictions of this model. Finally, we develop an extension of our modeling approach to quantitatively describe and experimentally evaluate RNA-templated and DNA-templated mechanisms for dsRNA formation.

**Results**

**Analysis of Initiation-Elongation Kinetic Model**

The structure of the initiation-elongation model used in this work is a variant of a model postulated (but not experimentally explored) in a past publication[17]. In this model, polymerase (P) and the DNA promoter ($DNA_p$) reversibly bind to form a complex ($P \cdot DNA_p$) that can undergo transcription initiation at a rate $k_i$. After initiation, the polymerase translocates across the DNA sequence in an elongation state ($P_E$) with an effective rate constant $k_{e,\text{tot}}$. Initiation frees the DNA promoter to be further bound by incoming polymerase. Transcription termination was assumed to be instantaneous for the linearized DNA templates used in RNA manufacturing.

In addition, this model neglects the formation of short sequences resulting from abortive transcription. Considering that these aborts comprise a negligible mass fraction of the IVT product of long transcripts, this abortion process can be considered part of the effective dissociation rate of the initiation complex. Schematically, the transcription process is represented as

$$\text{P} + \text{DNA}_p \underset{k_{\text{off}}}{\overset{k_{\text{on}}}{\rightleftarrows}} \text{P} \cdot \text{DNA}_p \underset{-\text{DNA}_p}{\overset{k_i}{\rightarrow}} \text{P}_E \underset{-\text{P}}{\overset{k_{e,\text{tot}}}{\rightarrow}} \text{RNA}. \quad (1)$$

This reaction network was approximated to operate in a quasi-steady state as the time associated with the synthesis of a single transcript as measured by single-molecule experiments (3–30 s) is substantially lower than the time constant of substrate consumption for the data in this work (0.25–0.5 hr)[10,18]. Each of the kinetic processes in this model are first-order approximations of multi-step phenomena that are dependent on solution conditions, notably temperature, pH, nucleoside triphosphate (NTP), and Mg concentrations.

Here we model the IVT reaction rate as a function of DNA and RNA polymerase concentrations to serve as a framework for understanding the effects of other process variables. The rate of RNA synthesis is equivalent to the initiation rate,

$$R_{\text{tr}} = k_i [\text{P} \cdot \text{DNA}_p], \quad (2)$$

which is dependent on the concentration of the initiation complex. This complex concentration is derived using a quasi-steady state approximation (SI Section 1),

$$[\text{P} \cdot \text{DNA}_p] = \frac{[\text{P}]_{\text{tot}} + \alpha[\text{DNA}]_{\text{tot}} + K_{\text{MD}} - \sqrt{([\text{P}]_{\text{tot}} + \alpha[\text{DNA}]_{\text{tot}} + K_{\text{MD}})^2 - 4\alpha[\text{P}]_{\text{tot}}[\text{DNA}]_{\text{tot}}}}{2\alpha}, \quad (3)$$

where $[\text{P}]_{\text{tot}}$ and $[\text{DNA}]_{\text{tot}}$ are total polymerase and DNA concentrations and

$$\alpha = 1 + \frac{k_i}{k_{e,\text{tot}}}, \quad K_{\text{MD}} = \frac{k_i + k_{\text{off}}}{k_{\text{on}}}. \quad (4)$$

This hypothesized model differs from the approach of Arnold et al.[12] in two key ways. Firstly, the removal of DNA promoter and RNA polymerase at different points in the process allow for a single DNA chain to feature multiple bound elongating polymerase molecules. Secondly, no assumptions are made regarding the relative concentration of DNA and polymerase during derivation of the rate law, which allows the model to operate across a broader space of species concentrations. For $\alpha = 1$, the proposed model converges to the structure of the rate law used by Martin and Coleman[19] in describing oligonucleotide transcription rates, which is based on an assumption that the effect of elongation is negligible. This rate law is referred to as an *initiation-*

*limited* model in this work and is a special case of the presented initiation-elongation model.

The initiation-elongation model predicts that the limiting step depends on relative DNA and polymerase concentrations. In the regime where $\alpha[\text{DNA}]_{\text{tot}} \ll [\text{P}]_{\text{tot}}$, the limiting factor is the number of DNA promoter binding sites, and the overall rate converges to

$$R_{\text{tr}} = \frac{k_i[\text{DNA}]_{\text{tot}}[\text{P}]_{\text{tot}}}{[\text{P}]_{\text{tot}} + K_{\text{MD}}}. \tag{5}$$

In this limit, the predictions of the initiation-limited model and the initiation-elongation model converge. We refer to this limit as the *initiation-limited regime*. Conversely, in the regime where $[\text{P}]_{\text{tot}} \ll \alpha[\text{DNA}]_{\text{tot}}$, the rate law converges to

$$R_{\text{tr}} = \left(k_i^{-1} + k_{e,\text{tot}}^{-1}\right)^{-1} \frac{[\text{DNA}]_{\text{tot}}[\text{P}]_{\text{tot}}}{[\text{DNA}]_{\text{tot}} + \frac{K_{\text{MD}}}{\alpha}}, \tag{6}$$

in which the rate of transcription is limited by the combined timescale of initiation and elongation. Considering that for long (>1000 base pair) sequences, $k_{e,\text{tot}} < k_i$, we refer to this case as an *elongation-limited* regime.

**Initiation-elongation model is necessary to describe IVT kinetic data**

To validate the structure of the initiation-elongation model, the rate of transcription of a 2078 base pair DNA template encoding the firefly luciferase gene (Fluc) was measured as a function of DNA and T7 RNA polymerase concentrations (Figure 1). Both the initiation-limited model (with parameters $K_{\text{MD}}$ and $k_i$) and the initiation-elongation model (with parameters $K_{\text{MD}}$, $k_i$, and $\alpha$) are fit to these data. As the solution concentrations of DNA and T7 RNA polymerase used in these experiments are too high to give identifiable estimates for $K_{\text{MD}}$, a Bayesian prior for $K_{\text{MD}}$ of 50±25 nM was used based on previous measurements[3]. This was acceptable for the fitting process as the main goal of these experiments was to estimate $k_i$ and $\alpha$. Bayesian information criterion analysis showed that the additional parameter of the initiation-elongation model provided a significant improvement in fitting over the initiation-limited model (SI Section 2.5). The initiation-elongation model (unlike the initiation-limited model) describes key trends in the data, including the linear relationship between RNA polymerase concentrations and reaction rate in the high-DNA regime and the linear relationship between DNA template concentration and reaction rate in the low-DNA regime. Uncertainty analysis indicates that the parameter estimates of the initiation-elongation model are practically identifiable and that the uncertainty region of $k_i$ and $\alpha$ are not highly correlated with uncertainty in $K_{\text{MD}}$. This indicates that the exact choice for the prior value of $K_{\text{MD}}$ has a minor

effect on the estimates of $k_i$ and $\alpha$.

With the structure of the initiation-elongation model validated, a model-based design of experiments (MBDOE) approach using the D-optimal criterion was employed to choose experimental conditions that best identify the two key kinetic parameters ($k_i$ and $\alpha$) of three more DNA sequences with differing DNA sequence length and initiation sequence: a dodecamer sequence matching the first 12 base pairs of the Fluc sequence and sequences coding for the COVID spike protein and EGFP protein (Table 1). MBDOE analysis indicated that two experiments corresponding to the initiation and elongation limited regimes were sufficient to practically identify the two parameters. When necessary to achieve greater parameter precision after one round of data collection, the MBDOE process was iterated. Table 1 shows the sequences used, their length, first three initiating base pairs, and their estimated kinetic parameters. An average per-base pair elongation rate constant is calculated as

$$k_{e,\text{bp}} = N_{\text{all}} k_{e,\text{tot}} \tag{7}$$

to aid in comparison between sequences.

Table 1: Transcription parameters of DNA sequences. Each sequence is characterized by its length and first three initiating nucleotides (init).

|  |  |  | Parameter | | | |
| --- | --- | --- | --- | --- | --- | --- |
|  | Length | Init. | $k_i$ (h$^{-1}$) | $\alpha$ | $k_{e,\text{tot}}$ (h$^{-1}$) | $k_{e,\text{bp}} \times 10^{-5}$ (h$^{-1}$) |
| Fluc | 2078 | AGA | 1220±100 | 9.9±0.7 | 136±7 | 2.8±0.2 |
| Fluc dodecamer | 12 | AGA | 1500±300 | 0.7±0.3 | – | – |
| COVID | 4243 | AGA | 1280±150 | 20±4 | 70±10 | 2.9±0.4 |
| EGFP | 942 | GGG | 2800±500 | 15±3 | 200±30 | 1.9±0.3 |

**Kinetic modeling indicates that polymerase-polymerase interactions and pausing during elongation have nonzero but unidentifiable effects on macroscopic reaction rates**

The initiation-elongation model is a minimal approach to understanding the kinetic trends of the IVT system and uses a number of approximations. One key approximation is that all polymerase molecules in the elongation state advance with the same rate constant regardless of their position on the chain or the local density of polymerase molecules. This approximation is not valid in the case where polymerase molecules can hinder each other's progress along the DNA sequence[20]. In addition, this polymerase-polymerase exclusion can be exacerbated by the pausing of polymerase during elongation[21,22]. The possibility of polymerase-polymerase interactions and pausing raises several questions relevant for the engineering of the IVT system. In what regimes, if any, can these polymerase-polymerase interactions distort the predictions of the initiation-elongation model presented above? In addition, can the extent of these interactions be assessed using macroscopic rate measurements?

To answer these questions, we developed a kinetic model that extends the initiation-elongation model to represent elongation as a totally asymmetric simple exclusion process (TASEP). The model assumes that polymerase molecules elongate by unidirectional transitions between $M$ $L$-nucleotide sized segments, where $L$ is the estimated exclusion width of the polymerase molecule. Schematically, this model has the structure

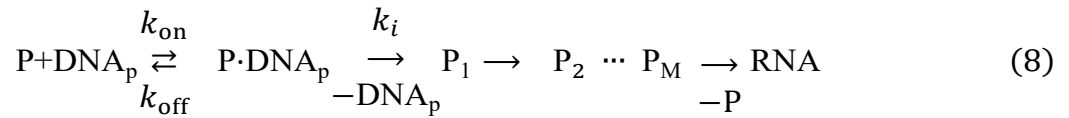

$$\text{P} + \text{DNA}_p \underset{k_{\text{off}}}{\overset{k_{\text{on}}}{\rightleftarrows}} \text{P} \cdot \text{DNA}_p \xrightarrow[-\text{DNA}_p]{k_i} \text{P}_1 \rightarrow \text{P}_2 \cdots \text{P}_M \xrightarrow[-\text{P}]{} \text{RNA} \tag{8}$$

where $P_i$ represents a polymerase molecule on the $i$th segment. Exclusion between particles is represented by the form of the rate law describing translocation between segments. Past work has focused on developing and validating mean-field approximations for these rates that account for both polymerase-polymerase interactions and pausing of polymerase molecules during elongation[22]. Using these rate laws, we developed an approximate analytical expression to predict macroscopic reaction rates (SI Section 2). In the most general case including pausing and polymerase-polymerase interactions, the derived rate law is

$$R_{\text{tr}} = k_i[\text{P·DNA}_\text{p}] \frac{[\text{DNA}]_{\text{tot}} - \frac{\beta[\text{P·DNA}_\text{p}]}{K_{MD} - \theta}}{[\text{DNA}]_{\text{tot}} + \gamma[\text{P·DNA}_\text{p}]} \tag{9}$$

where the initiation complex $[\text{P·DNA}_\text{p}]$ is defined by

$$\alpha\gamma \left(\frac{[\text{P·DNA}_\text{p}]}{[\text{DNA}]_{\text{tot}}}\right)^3 + \left(\alpha + \frac{\beta}{[\text{DNA}]_{\text{tot}}} - \gamma\left(\frac{[\text{P}]_{\text{tot}}}{[\text{DNA}]_{\text{tot}}} + \frac{\theta}{[\text{DNA}]_{\text{tot}}} + \alpha\right)\right)\left(\frac{[\text{P·DNA}_\text{p}]}{[\text{DNA}]_{\text{tot}}}\right)^2$$

$$- \left(\frac{[\text{P}]_{\text{tot}}}{[\text{DNA}]_{\text{tot}}} + \alpha + \frac{K_{MD}}{[\text{DNA}]_{\text{tot}}} - \gamma\frac{[\text{P}]_{\text{tot}}}{[\text{DNA}]_{\text{tot}}}\right)\left(\frac{[\text{P·DNA}_\text{p}]}{[\text{DNA}]_{\text{tot}}}\right) + \frac{[\text{P}]_{\text{tot}}}{[\text{DNA}]_{\text{tot}}}$$

$$= 0 \tag{10}$$

with the composite parameters

$$\gamma = k_i \frac{f\tau^2}{1 + f\tau}, \quad \beta = \frac{Lk_i^2}{k_e k_{\text{on}}}, \quad \theta = \frac{k_{\text{off}}}{k_{\text{on}}}, \tag{11}$$

where $f$ and $\tau$ represent the frequency of pausing events and the timescale of pauses, respectively, following the approach of Wang et al.[22] We refer to the model in this general case as the *long pause* (LP) model. Intuitively, the dimensionless parameter $\gamma$ represents the relative importance of pausing to the transcription system, and setting $\gamma$ to zero results in a simpler model that neglects the effect of pauses (which we call the *short pause* (SP) model). The parameter $\beta$ represents the relative importance of polymerase-polymerase interactions on the system. Setting $\beta$ to zero recovers the initiation-elongation model.

The complexity and number of parameters of the LP model raises a question of practical identifiability. If the transcription system truly behaved in accordance with equations 9 and 10, could it be distinguished from the initiation-elongation model with macroscopic measurements? Can macroscopic measurements be used to identify the parameters $\gamma$ and $\beta$? Bayesian information criterion analysis indicates that neither of the SP or LP models fit the kinetic data collected in this work better than the initiation-elongation model (SI Section 2.5). Moreover, the same is true for synthetic data generated by the SP and LP models in the case of reasonable estimates for microscopic parameters and experimental noise. In fitting this synthetic data, neither the SP nor LP model can identifiably recover estimates for all of their kinetic parameters, resulting in highly correlated parameter confidence regions.

While the initiation-elongation model can describe the output of the more complicated LP and SP models, the resulting fitted kinetic parameters do not match the microscopic

ground truth values used to generate the data. This implies that the measured kinetic parameters in Table 1 may serve as effective parameters that do not perfectly reflect microscopic rates of initiation and elongation. The effective initiation rate constant estimated from data generated by the SP model is approximately 80–90% of the ground truth value. In the case of data generated by the LP model, the estimated initiation rate constant is dependent on values of $\gamma$ but can be significantly lower (SI Section 2.5). The estimated elongation rate constant is not distorted by more than 10% in either of these cases.

**Initiation-elongation model describes sensitivity of transcription rates to salt addition**

Past research has noted that the addition of salts, including the necessary magnesium cofactor, to the IVT system can decrease transcription rates by disrupting the binding between RNA polymerase and the DNA promoter[3,23]. The initiation-elongation model is a useful tool for quantitative understanding the effect of salt concentrations on transcription rates. Equations 5 and 6 show a non-obvious emergent result of the initiation-elongation model. While both initiation- and elongation-limited regimes can be described using a Michalis-Menten structure, the effective Michaelis-Menten constant differs between the two by a factor of $\alpha$. Intuitively, the elongation-limited regime is less sensitive than the initiation-limited regime to disruptions in polymerase-promoter binding.

In order to describe the effect of salt addition on overall transcription rates, a model for the effect of salt concentrations on $K_{MD}$ is required. The IVT system contains multiple salts, including NTPs, magnesium, buffers, and associated counterions. It has been shown that different salts affect transcription rates to different degrees[23]. While the literature on salt effects on protein-DNA binding is substantial[24], there is little published work on the practical problem of modeling this relationship in the context of mixed-salt systems relevant to IVT. We adapted a previously proposed approach that augments the predictions of counterion condensation (CC) theory with an effective salt concentration[25]:

$$K = \frac{k_{\text{off}}}{k_{\text{on}}} = K_0 \left(\frac{[\text{salt}]}{1 \text{ mole}}\right)^n \quad (12)$$

where $K_0$ represents the intrinsic binding strength of the DNA promoter, and [salt] is an effective salt concentration calculated as

$$[\text{salt}] = \sum_{i=1}^{N_{\text{ion}}} \omega_{\text{ion},i} [\text{ion}_i] \quad (13)$$

where $\omega_{\text{ion},i}$ is a weighting factor specific to each cation and anion $i$ in the IVT system (SI Section 3.2). While this relation was developed as an empirical extension of classical CC theory[26], we show that it emerges naturally from an extended treatment of CC theory that includes the presence of multiple salts (SI Section 3). While CC theory is a dramatically simplified rendering of the physics of ligand-DNA interactions that can be modeled with greater fidelity using computational techniques that incorporate biomolecule structure and diffuse ion binding[27,28], this relation describes key trends in experimental binding data of mixed salt systems and is useful in an engineering context. To adapt this thermodynamic relation to kinetic modeling, we assume that $k_{\text{on}}$ is constant at a value of $204\,\text{h}^{-1}\,\text{nM}^{-1}$.[18] While $k_{\text{on}}$ has been shown to be sensitive to salt[25], this approximation is valid in the high salt limit in which $K_{\text{MD}}$ is very large. According to CC theory, the constant $n$ represents the number of cations displaced upon polymerase-DNA binding, a prediction which has been shown to correlate with experimental results[29]. Based on an approximate structural analysis of T7 RNA polymerase, we hypothesized that a reasonable estimate for $n$ was $5.0$.[8] We validated this choice of $n$, fitting a single parameter $K_0$, on published data of $K_{\text{MD}}$ as a function of sodium chloride addition (Figure 2A)[3]. We find that this model suitably describes changes in $K_{\text{MD}}$ in the high-salt limit.

To test the prediction of the initiation-elongation model that the salt sensitivity of the IVT reaction differs between the initiation and elongation limited regimes, we measured the transcription rate of the Fluc construct as a function of sodium chloride addition. Two reaction schemes with different DNA concentrations (117.5 and 9.2 nM) are tested with solution conditions (including 192 nM T7 RNA polymerase) otherwise held constant. The ratios $\alpha[\text{DNA}]_{\text{tot}}/[\text{P}]_{\text{tot}}$ of the two schemes are 5.6 and 0.5, reflecting that these experiments probed an elongation-limited regime and a region primarily governed by initiation limitation, respectively. The two reaction schemes exhibited significantly different responses to salt addition (Figure 2B). After fitting a single parameter $K_0$ describing the general binding strength of the Fluc promoter, the initiation-elongation model described these trends. The difference in behavior between reaction conditions cannot be predicted a model which only considers initiation limitations.

In order to understand how salt sensitivity varies between sequences, the same kinetic measurements are performed on the EGFP sequence using an IVT scheme analogous to the low-DNA Fluc kinetics discussed above (containing equal DNA concentrations by

mass of transcribed region). If the parameter $K_0$ of the two sequences is the same, model prediction indicates that the salt sensitivity should be the same as the Fluc construct within the range of experimental precision. It was instead found that the EGFP sequence was significantly less sensitive to salt addition than the COVID sequence. In our modeling framework, this lower sensitivity is parametrized as a lower value for $K_0$, that is, stronger intrinsic binding strength. Table 2 shows estimated values of $K_0$ for the two tested sequences, as well as the implied $K_{MD}$ in the absence of sodium chloride addition. These calculated values of $K_{MD}$ validate the Bayesian prior used for the estimation of parameters in Table 1.

Table 2: Estimated binding parameters of measured sequences. Each sequence is characterized by its length and first three initiating nucleotides (init).

|      | Parameter | | | |
| --- | --- | --- | --- | --- |
|      | **Length** | **Init.** | $K_0$ **(mM)** | $K_{MD}$ **at [NaCl] added = 0 (nM)** |
| Fluc | 2078 | AGA | 1.5–3.0 | 40–70 |
| EGFP | 942  | GGG | 0.5–1.0 | 25–35 |

**Extending modeling approach to the formation of dsRNA impurities**

Double-stranded RNA (dsRNA) is an immunogenic byproduct of the IVT reaction. These dsRNA byproducts are heterogeneous in size and sequence, and a given RNA product molecule may contain both single and double stranded regions. As such, dsRNA is challenging to remove in downstream processing steps of RNA manufacturing. Two mechanisms for dsRNA formation in IVT have been proposed, which share undesired polymerase binding as a common feature. A mechanism of RNA self-templated extension has been shown to produce short double-stranded segments in oligomeric model systems[15], and has been used as the conceptual basis for strategies to decrease dsRNA formation based on immobilization and high salt concentrations[30]. Conversely, a mechanism of DNA-templated antisense RNA synthesis has been shown to form hybridized dsRNA structures for specific sequences[16]. While each of these mechanisms has been used as the conceptual basis for engineering strategies to reduce dsRNA formation, there is no work in understanding the quantitative implications of these models for

input-output relations of dsRNA formation.

Considering that undesired RNA polymerase binding is the foundation of both mechanisms, the modeling approach developed in this work, which explicitly considers both free polymerase solution concentrations and polymerase-DNA binding, serves as a necessary platform for modeling dsRNA formation kinetics. We developed an extension of the initiation-elongation model to incorporate the binding of free RNA polymerase to either an undesired promoter on RNA ($RNA_{up}$) or an undesired promoter on the antisense DNA strand ($DNA_{up}$) (Figure 3A). This model considers dsRNA to be a homogeneous chemical species. While this description is not perfectly representative of the known heterogeneity of dsRNA, it is appropriate for understanding trends in macroscopic dsRNA quantities. Using this schematic model, we derived quantitative input-output relations to model the fraction of dsRNA in the IVT product (SI Section 4). In addition to the approximations used to derive the initiation-elongation model, we assume that the amount of dsRNA product is much less than the ssRNA product and that undesired binding is a relatively rare event compared to the desired binding.

For a mechanism of RNA-templated dsRNA formation, our modeling approach predicts that the product dsRNA fraction is proportional to

$$\frac{[dsRNA]_t}{[R]_t} \propto \frac{[R]_t}{[DNA]_{tot} - [P \cdot DNA_p]} \tag{14}$$

where $[dsRNA]_t$ and $[R]_t$ are the concentrations of dsRNA and total RNA at a given extent of reaction, and $[P \cdot DNA_p]$ is the same quantity given by equation 3. In the case of DNA-templated antisense synthesis, generating mechanistic predictions is more difficult given the dynamics of the sense-antisense hybridization step. Kinetic studies of analogous systems indicate that the rate constant of this hybridization is $10^{-5}$–$10^{-4}$ min$^{-1}$ nm$^{-1}$, which implies a time constant of approximately 1–10 minutes for the reaction concentrations used in this work. Considering that the time constant of the IVT reactions studied in this work take place on time scales of about 20–600 minutes, the hybridization step was approximated as instantaneous. While this approximation may neglect these hybridization trends, it lends a dramatic simplification to model predictions. In the case of DNA-templated dsRNA formation, our modeling approach predicts that

$$\frac{[dsRNA]_f}{[R]_f} \propto \frac{[DNA]_{tot}}{[DNA]_{tot} - [IC]} \tag{15}$$

Equations 14 and 15 can either be viewed as competing models, or as two components of a

larger modeling strategy that includes both RNA- and DNA-templated pathways. In the context of this work, we focus on their evaluation as competing models.

The macroscopic predictions of these models differ in two key ways. First, the RNA-templated model predicts that the dsRNA fraction (the ratio of dsRNA concentration to total RNA concentration) is low at early timepoints and rises linearly with respect to reaction conversion, while the DNA-templated model predicts that the dsRNA fraction is constant with respect to reaction conversion. In addition, the RNA-templated model, which assumes a competition between RNA and DNA as binding sites, predicts that dsRNA formation should trend to zero as the concentration of DNA is increased. The DNA-templated model predicts some dependence between input DNA and dsRNA formation, but predicts a finite asymptotic value of dsRNA formation.

To evaluate these models, we measured the final mass fraction of dsRNA in the Fluc IVT product as a function of the extent of reaction (Figure 3B). Above a fractional conversion of 40%, the dsRNA fraction of the system was relatively constant, in line with the DNA-templated model. However, timepoints collected at earlier conversions showed a decreasing trend, which is not consistent with the predictions of either model. The same measurements performed on the COVID and EGFP constructs showed a similar nonincreasing result (SI Section 5). In addition, we measured the dsRNA fraction at complete conversion as a function of input DNA (Figure 3C). We found that these data did not exhibit a clear statistical trend, but is dramatically inconsistent with the predictions of the RNA-templated model. We found that varying the concentrations of polymerase enzyme and salts did not give a significant effect on final dsRNA fractions (SI Section 5).

**Discussion**

Given the ubiquity of in vitro transcription in industrial RNA manufacturing, a kinetic framework that incorporates both DNA template and RNA polymerase concentrations to predict rates of RNA synthesis is a crucial tool to effectively use both expensive catalysts. In this work, we find that a model which incorporates both initiation and elongation steps is required to describe kinetic data across a range of DNA and RNA polymerase concentrations. Contrary to previous reports, we show that the limiting step is dependent on solution conditions and that the reaction system can be limited by the rate of elongation.

A primary goal of this work is to inform rapid and economical IVT process development, which currently involves data-driven designs of experiments to assess input-output relations.

Since different DNA sequences transcribe with different kinetics, these designs are often repeated for each manufactured sequence. In our modeling approach, differences between sequences are encoded by kinetic parameters. We demonstrate that the key kinetic parameters of transcription ($k_i$ and $\alpha$) can be identified with only two experiments (taken in the elongation and initiation limited regimes, respectively). The measured values of these parameters are correlated with physical intuition and prior literature (Table 1). Sequences with the same three initiating base pairs (AGA) had the same initiation rate constant within the uncertainty of our measurements (1100–1800 h$^{-1}$). The EGFP construct, which contained the canonical initiation sequence (GGG), exhibited a significantly higher initiation rate constant (2300–3300 h$^{-1}$). These values are in the general range of previous reports from both single molecule and oligomeric studies (1000–2000 h$^{-1}$)[3,18] and are consistent with reports that mutations to the canonical initiation sequence decreased overall transcription rates[31]. We find that the parameter $\alpha$, which describes the relative importance of elongation in the transcription process, is loosely correlated with sequence length. This in turn implies that the effective per-base-pair elongation rate is within 1.6–3.0×10$^5$ h$^{-1}$ for all sequences tested. This is in the same order of magnitude as previously reported values from single molecule studies (4–8×10$^5$ h$^{-1}$)[11,32]. In addition, the estimated $\alpha$ of a tested oligomeric sequence is 1.0 within the interval of uncertainty, which confirms our intuition that transcription of oligomers is purely initiation limited.

Differences between the rate constants calculated in this work and those reported by previous researchers can be explained by two causes. These kinetic parameters are first-order approximations of multiple steps and as such are dependent on NTP concentrations, pH, temperature, and other solution conditions. Solution concentrations of NTPs, magnesium, and other salts are typically much higher in the context of industrial RNA manufacturing (and this work) than in most fundamental studies of transcription kinetics. Secondly, kinetic modeling indicates that polymerase-polymerase interactions during elongation can manifest as a decrease in the effective initiation rate constant when analyzed with the initiation-elongation model. If the estimated initiation rate constant of a DNA sequence is significantly less that the initiation rate constant of its initiating oligomer sequence, these polymerase-polymerase interactions could be a cause. We do not observe a significant difference between the Fluc and Fluc dodecamer sequence in this work, however.

Recent trends in IVT reaction engineering have added new relevance to the effect of salts, which disrupt polymerase-promoter binding, on transcription rates. Salt addition has been

proposed as a method to shift transcription away from dsRNA impurities, which introduces tradeoffs in the context of RNA manufacturing[30]. In addition, industrial IVT schemes, including fed-batch reactions, increasingly use high NTP and Mg concentrations, which increases salt concentrations. Intuitively, transcription in the elongation-limited regime should be less affected by binding disruptions than transcription in the initiation-limited regime. We find that a simple semi-empirical model can describe trends in $K_{MD}$ as a function of salt concentration (Figure 2A). Using this relation in combination with the initiation-elongation model, we predict the experimental result that transcription in the high-DNA elongation-limited regime is much less sensitive to salt addition than transcription in the low-DNA initiation-limited regime (Figure 2B). Understanding this difference in sensitivity can inform reaction design in the context of industrial RNA manufacturing. In addition, we find that salt sensitivity varies between the Fluc and EGFP constructs studied in this work, implying that the two sequences differ in their inherent polymerase binding strength (Figure 2C, Table 2). We hypothesize that this difference is due to a difference in the initiation sequence: while the EGFP construct features the canonical sequence, the Fluc construct uses a modified sequence to accommodate a co-transcriptional AG capping agent (eg. CleanCap). This difference in binding may be related to the difference in estimated initiation rate constants between the two sequences by a common mechanistic root. This finding illustrates a tradeoff in the use of AG analogues for co-transcriptional capping: strategies involving high salt concentrations, such as fed-batch reactions, may need to be significantly modified to accommodate these capping technologies.

A key application of kinetic modeling in the context of IVT is in understanding input-output relationships for impurity formation. In this work, we extend the initiation-elongation model to consider two proposed mechanisms for formation of double-stranded RNA (dsRNA): 3' RNA self-templated transcription and DNA-templated antisense transcription followed by hybridization (Figure 3A). We evaluate the predictions of these models relative to a macroscopic binding assay that estimated the total concentration of dsRNA in the system. While these assays typically cannot detect small (~40 base pair) regions of dsRNA, they have been shown to correlate with in vivo immune response[33]. As such, we regard them as an effective measurement of the class of dsRNA (i.e., long dsRNA) that is of interest in the manufacturing process.

We find that the RNA-templated mechanism for dsRNA formation was extremely unlikely to be the source of our experimental data. The dsRNA fraction in our reaction system was

constant or decreasing as a function of reaction conversion, in contrast to the prediction of the RNA-templated model that the dsRNA fraction should increase as more RNA is synthesized (Figure 3B). The RNA-templated model predicts that adding more DNA to the IVT system should shift the kinetic competition for polymerase molecules away from RNA and proportionally decrease dsRNA formation. In contrast to these predictions, we find that increasing DNA concentrations did not significantly affect final dsRNA fractions (Figure 3C). We hypothesize that the relative importance of the RNA 3' self-templated mechanism decreases with increasing template length, as the relative concentration of the 3' end decreases, and that oligomeric model systems may overstate the relevance of this mechanism for the formation of the long dsRNA of interest in RNA manufacturing.

The implication of our experimental data for the validity of the DNA-templated model is less clear. While the DNA-templated model does not diverge as dramatically from experimental results, it cannot describe the decrease in dsRNA fraction in early stages of the reaction. In addition, dsRNA fraction data collected by varying the concentration of polymerase enzyme and salt addition show ambiguous results which do not indicate that the predictions of the DNA-templated model are more effective than a constant null hypothesis (SI Section 5). A key approximation of our modeling approach is neglecting the kinetics of hybridization, which may be important for describing trends in these data. While our kinetic results did not show a significant effect of solution conditions on final dsRNA fractions, we found that these results could be dramatically influenced by small variations in sample processing after reaction quenching (the results shown in this work were generated by controlling processing conditions for all samples). This implies that processing plays a key role in the detection of dsRNA impurities, which may be related to hybridization, and that the results of dsRNA quantitation measurements cannot by fully understood by solely analyzing the IVT step.

While the presented results do not definitively identify a mechanism for explaining trends in macroscopic dsRNA formation, the modeling approach in this work serves as a platform for both future work and understanding trends in previously reported data. In both the RNA- and DNA-templated models, dsRNA formation is proportional to the ratio of Michaelis-Menten constants of desired and undesired promoter binding, respectively (SI Section 4). This implies that reaction engineering strategies which differentially impact these two bindings can be used to limit dsRNA formation. This modeling observation gives quantitative structure to an array of strategies that previous researchers (with either mechanism in mind) have used to decrease

dsRNA formation, including the use of engineered polymerase enzymes[34], high temperatures[35], salts[30], and chaotropic agents[36]. In addition, our modeling approach provides a mechanistic explanation for a previously reported result that replacing the initiating nucleotide (canonically G) with A increases dsRNA formation[33], as our previous results indicate that this base change is likely to decrease the binding strength of the desired initiation complex.

**Experimental Procedures**

**In Vitro Transcription Kinetic Measurements**

All in vitro reactions studied in this work took place and pH 8.0 and contained 5 mM of each NTP (ATP, CTP, GTP, and N1-Methylpseudouridine-5'-Triphosphate), 21.075 mM $MgCl_2$, 45 mM of pH 7.9 Tris-HCl buffer, 2 mM spermidine, 10 mM DTT, 6 U/mL of inorganic pyrophosphatase, and 400 U/mL of RNAase inhibitor. All reaction materials were acquired from Hongene Biotech, other than $MgCl_2$, which was acquired from Thermo Fischer. Transcription reactions were assembled at volumes between 50–100 µL and incubated at 37˚C. Aliquots of 6 µL were periodically removed and quenched in 60 µL of 50 mM EDTA. These quenched samples were further diluted 36-fold (for a total dilution of 400-fold) and analyzed with the HPLC method of Welbourne et al.[37] to quantify the concentrations of the four NTPs. Linear regression analysis was used to quantify the rate of NTP decay. While an orthogonal analysis of the RNA product was possible, it was found that quantification of NTPs was less sensitive to both systematic and random experimental errors. In order to ensure that data points represented the initial rate of reaction, points collected at high conversion (below 2 mM of the limiting NTP remaining) were excluded from this analysis.

**Measurement of dsRNA Concentrations**

The quantification of dsRNA was performed using the Lumit® dsRNA Detection Assay kit (Promega) according to the manufacturer's instructions. White 96-well plates were obtained from Thermo Fisher Scientific/Corning®. Diluted reaction samples were prepared as described in the previous method section. The diluted samples were subsequently mixed with the dsRNA assay buffer to achieve an expected final dsRNA concentration of 2 ng/mL per well. Three technical replicates of each reaction sample were measured using a Thermo Fisher Varioskan® Flash plate reader with an integration time of 500 ms, and the results were averaged. Experimental variance between these technical replicates was negligible relative to the variance between replicate reactions. Background luminescence was determined by averaging the readout from the 0 ng/mL dsRNA standard and was subtracted from all sample measurements.

To determine the dsRNA/mRNA fraction, mRNA concentration was quantified via HPLC using 400-fold diluted reaction samples.

**Parameter estimation and model-based design of experiments for kinetic model**

Model evaluation and parameter estimation are performed in the Julia language. In order to estimate the relevant kinetic parameters, the maximum likelihood estimate of the vector of parameters $p$ was

$$\min_p \left(y - u(p)\right)^\top V_y^{-1}(k)\left(y - u(p)\right) \tag{16}$$

where $y$ is the vector containing all of the experimental data used for estimating parameters and $u(p)$ is the vector of corresponding model outputs as a function of the parameter vector $p$. The error covariance matrix $V_{p^*}$ of the best-fit estimate $p^*$ is approximated by

$$\text{cov}(p^* - p_{\text{true}}) = V_{p^*} \approx \left(S^\top V_y^{-1} S\right)^{-1} \tag{17}$$

where $p_{\text{true}}$ denotes the true parameters and $S$ is the sensitivity of the model outputs with respect to the vector $p$.

To ensure precision in the estimated kinetic parameters, model-based design of experiments was performed to minimize the determinant of the estimated parameter covariance matrix, known as D-optimality. Given a hypothesized parameter set $\hat{p}$ and a prior covariance matrix $\text{cov}(\hat{p})$, experimental points $x$ were chosen by solving the optimization

$$\min_x \left|[S(x,\hat{p})^\top V_p^{-1} S(x,\hat{p})) + \text{cov}(\hat{p})^{-1}]^{-1}\right| \tag{18}$$

where $S(x, \hat{p})$ is the sensitivity matrix of the experimental points $x$ and the estimated parameters $\hat{p}$, and $|\cdot|$ is the determinant. For both optimizations (parameter estimation and design of experiments), the gradient-based L-BFGS optimization was performed using the ForwardDiff.jl and NLopt.jl packages in Julia to compute model output sensitivities to parameters and use those sensitivities in gradient-based optimizers, respectively.



**Figure 1. Transcription Kinetics of Fluc Sequence.**

Measured transcription rate as a function of T7 RNA polymerase and DNA concentrations with initiation-elongation model predictions after fitting. Other solution conditions are held constant as described in Methods section. Reaction rate is linear with respect to RNA polymerase in the regime of $\alpha[DNA]_{tot} \gg [P]_{tot}$ (points in upper left of graph) and linear with respect to DNA concentrations in regime of $\alpha[DNA]_{tot} \ll [P]_{tot}$ (lower right of graph). Error bars on data points represent estimated 1σ experimental error based on triplicate experiments. Shaded areas are 95% prediction intervals of model based on estimated covariance matrix.

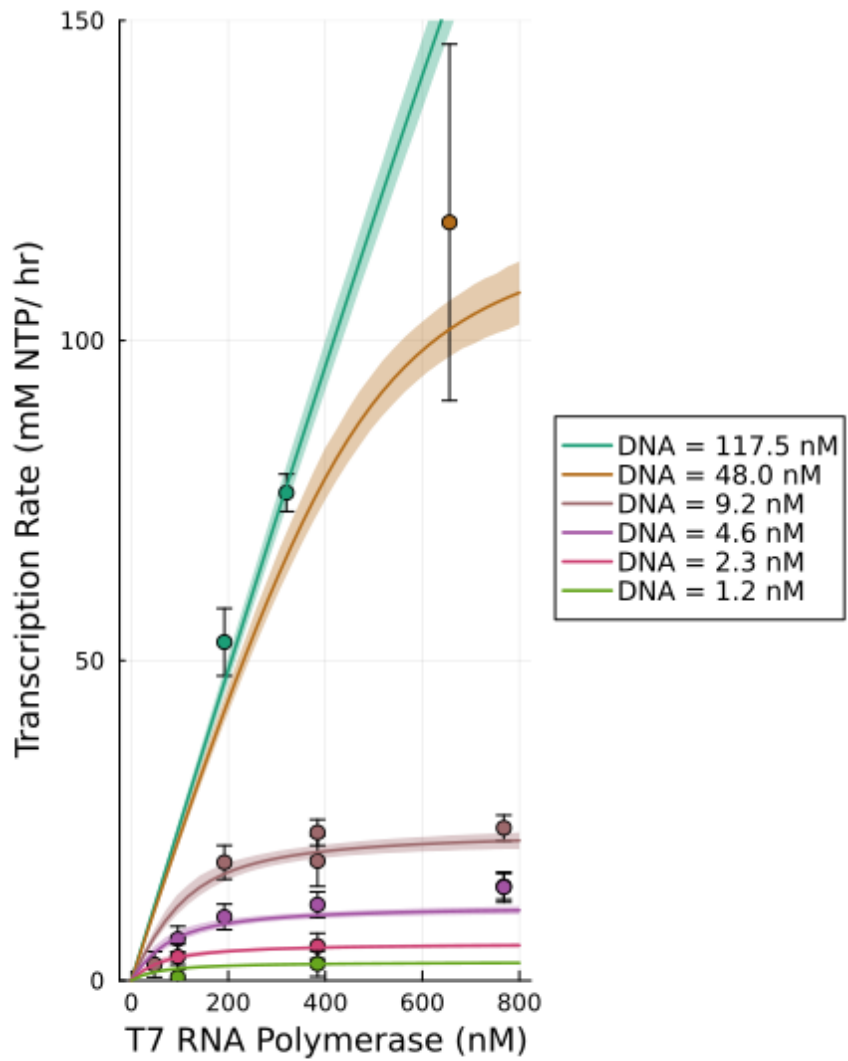

**Figure 2. Effect of salt addition on IVT kinetics.**

(A) Semi-empirical model describing effect of salt addition of polymerase-promoter binding can describe trends in $K_{MD}$, particularly in the range of high salt concentrations (>150 mM NaCl added). Data from Maslak and Martin[3]. (B) Applying this model for $K_{MD}$ to transcription kinetics of long sequences explains the difference in salt sensitivity between two reaction schemes that only differ in DNA concentration. The reaction scheme in the elongation-limited regime ([DNA]$_{tot}$ = 117.5 nM) is less sensitive that the scheme in the initiation-limited regime ([DNA]$_{tot}$ = 9.2 nM) to disruptions in polymerase-promoter binding due to salt addition. Model predictions are shown after fitting a single parameter $K_0$ (2.5 mM). (C) The EGFP sequence was measured to be less sensitive to salt addition than the Fluc sequence in analogous (equal DNA mass) reaction conditions. Model predictions are shown for best fit $K_0$ estimates for each sequence (0.75 mM for EGFP).

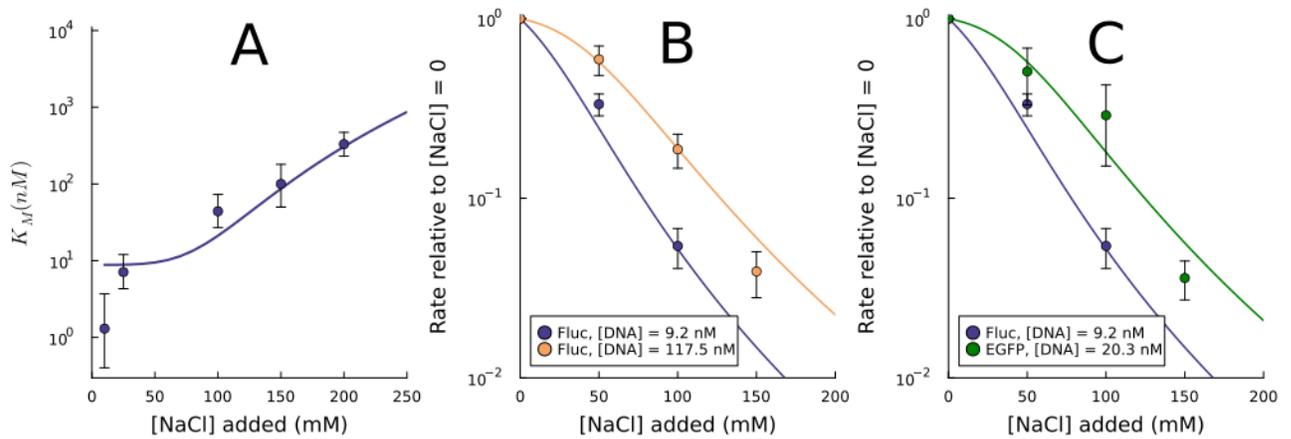

**Figure 3. Assessing Kinetic Models for dsRNA Formation.**

(A) Kinetic models for dsRNA formation are based on polymerase binding and initiation at undesired sites, which competes with the formation of the desired single-stranded product. In an RNA-templated mechanism, RNA polymerase binds to transient loop-back RNA structures and elongates across the RNA sequence, synthesizing dsRNA. In a DNA-templated mechanism, RNA polymerase binds to an undesired antisense promoter and synthesizes antisense RNA. These antisense RNA products hybridize with the main product RNA, which our model assumes is an instantaneous process. (B) In contrast to the predictions of the RNA templated model, we find that the fraction of dsRNA in the IVT product is constant or decreasing as a function of the extent of reaction. Model predictions are calibrated based on the final timepoint to show conceptual predictions. Reactions are performed with 192 nM of T7 RNA polymerase and 9.2 nM of Fluc DNA. (C) The dsRNA fraction after complete conversion is not significantly affected by DNA input concentrations. Model predictions are shown using the previous parameter calibration. Reactions are performed with 192 nM of T7 RNA polymerase.

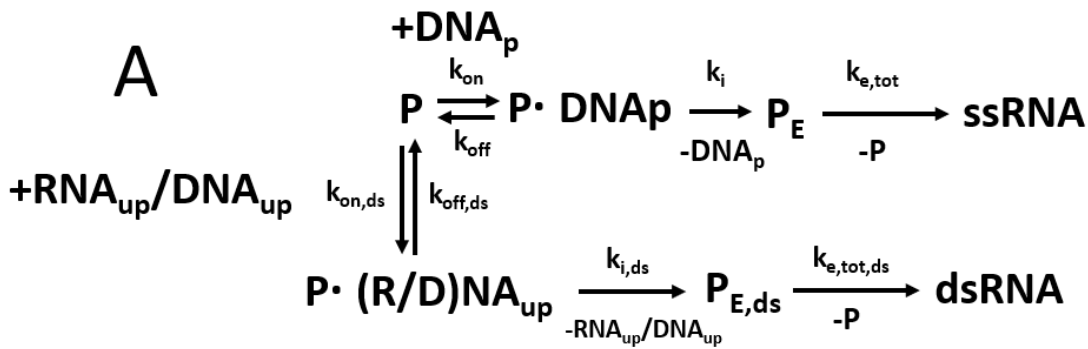

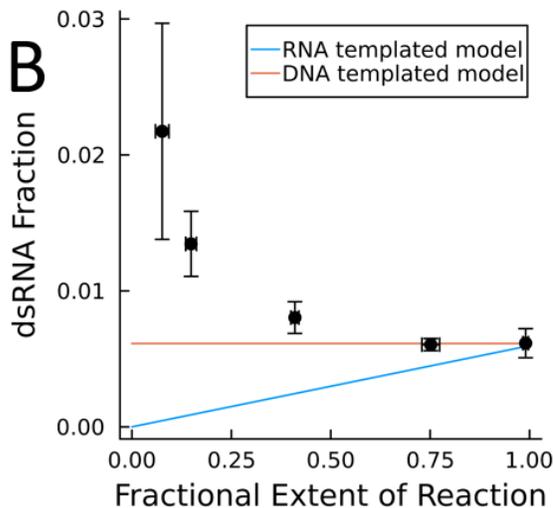
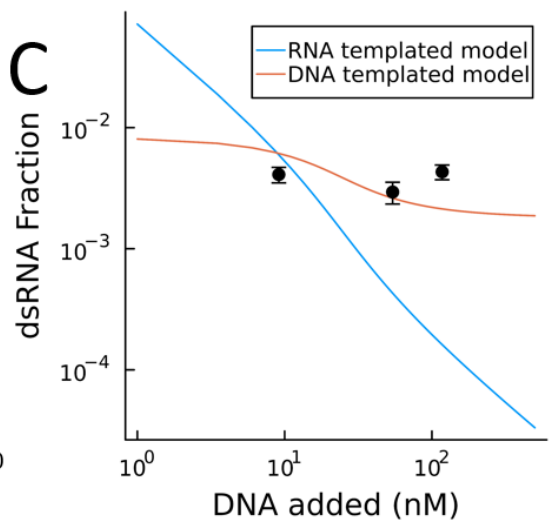


\maketitle
\tableofcontents
\section{Derivation of Initiation-Elongation Model}

The transcription process is modeled as containing three steps:
\begin{enumerate}
    \item The reversible binding of the DNA T7 promoter ($\mathrm{DNA_p}$) and T7 RNA polymerase $(\mathrm{P})$ to form an initiation complex $(\mathrm{P \cdot DNA_p})$.
    \item An irreversible initiation step which leads to polymerase elongating across the DNA chain ($\mathrm{P_{E}}$).
    \item The irreversible elongation of polymerase across the chain
\end{enumerate}
Schematically, this transcription process is
\begin{equation}\mathrm{P + DNA_p} \mathrel{\mathop{\rightleftarrows}^{k_\mathrm{on}}_{k_\mathrm{off}} } 
\mathrm{P \cdot DNA_p}\xrightarrow
[-\mathrm{DNA}_p]{k_{i}}\mathrm{P_{E}}\xrightarrow[-\mathrm{P}]
{k_{e\mathrm{,tot}}}\mathrm{RNA}\end{equation}
%\textcolor{black}
{where each step is assumed to be first order in the concentration of the associated macromolecules, P represents T7 RNA polymerase, and $\mathrm{DNA_p}$ represents the T7 promoter. All reaction rates are first-order using the rate constants written above. This formulation allows a DNA promoter to be available for reaction after initiation events, while the polymerase enzyme remains tethered to the DNA chain in the form of $\mathrm{P_{E}}$ until the elongation stage is completed. The reaction rate of interest is the rate of chain initiation,
\begin{equation}
V_\mathrm{tr} = \textcolor{black}{k_{i}\mathrm{[P \cdot DNA_p]}.}
\label{eq:two}
\end{equation}
The overall rate of the reaction is the initiation rate 
\begin{equation}
    R_\mathrm{tr} = k_i\mathrm{[P \cdot DNA_p]}.
\end{equation}
The concentrations of $\mathrm{P \cdot DNA_p}$ and $\mathrm{P_{E}}$ are both approximated as quasisteady. From the former, 
\begin{equation}
\mathrm{[P]}\mathrm{[DNA_p]} = K_\mathrm{MD}\mathrm{[P \cdot DNA_p]},
\end{equation}
where $\mathrm{[P]}$ and $\mathrm{[DNA_p]}$ are the free polymerase and DNA promoter concentrations, respectively, and
\begin{equation}
K_\mathrm{MD} = \frac{k_{i}+k_\mathrm{off}}{k_\mathrm{on}}.
\end{equation}

The concentration of the elongation state can be related to the concentration of the initiation complex using the quasisteady  assumption,
\begin{equation}
\textcolor{black}{\mathrm{[P_E]} = \frac{k_i}{k_{e\mathrm{,tot}}}}\mathrm{[P \cdot DNA_p]}.
\end{equation}
\textcolor{black}{With the dimensionless constant defined by}
\begin{equation}
\textcolor{black}{\alpha = 1+  \frac{k_i}{k_{e\mathrm{,tot}}}},
\end{equation}
the expression can be written as
\begin{equation}
    \textcolor{black}{\mathrm{[P_{E}]} = 
    (\alpha-1)\mathrm{[P \cdot DNA_p]}.}
\end{equation}

In this scheme, no assumptions are made about the relative concentration of the polymerase enzyme or DNA. The mass balances for each specie are
\begin{equation}\mathrm{[P]}_\mathrm{tot} = \mathrm{[P]}+\mathrm{[P \cdot DNA_p]}+\mathrm{[P_{E}]},\end{equation}
\begin{equation}\mathrm{[DNA_p]}_\mathrm{tot} = \mathrm{[DNA_p]}+\mathrm{[P \cdot DNA_p]}.\end{equation}
Using the quasisteady relations above, these expressions can be written as
\begin{equation}\mathrm{[P]}_\mathrm{tot} = \frac{K_\mathrm{MD}\mathrm{[P \cdot DNA_p]}}{{\mathrm{[DNA_p]}}}+\mathrm{[P \cdot DNA_p]}+(\alpha-1)\mathrm{[P \cdot DNA_p]},\end{equation}
\begin{equation}\mathrm{[DNA_p]} = \mathrm{[DNA_p]}_\mathrm{tot} - \mathrm{[P \cdot DNA_p]}.\end{equation}
Substituting the second equation into first gives
\begin{equation}\mathrm{[P]}_\mathrm{tot} = \frac{K_\mathrm{MD}\mathrm{[P \cdot DNA_p]}}{{\mathrm{[DNA_p]}_\mathrm{tot} - \mathrm{[P \cdot DNA_p]}}}+\alpha\mathrm{[P \cdot DNA_p]}.\end{equation}
Multiplying by the denominator gives
\begin{multline}
\mathrm{[P]}_\mathrm{tot}(\mathrm{[DNA_p]}_\mathrm{tot} - \mathrm{[P \cdot DNA_p]}) = 
K_\mathrm{MD}\mathrm{[P \cdot DNA_p]}+\textcolor{black}{\alpha}\mathrm{[P \cdot DNA_p]}(\mathrm{[DNA_p]}_\mathrm{tot} - \mathrm{[P \cdot DNA_p]}).
\end{multline}
Rearranging the expression gives
\begin{multline}
\mathrm{[P]}_\mathrm{tot}\mathrm{[DNA_p]}_\mathrm{tot}-\mathrm{[P]}_\mathrm{tot}\mathrm{[P \cdot DNA_p]} = K_\mathrm{MD}\mathrm{[P \cdot DNA_p]}-\textcolor{black}{\alpha}\mathrm{[P \cdot DNA_p]}^2\\+\textcolor{black}{\alpha}\mathrm{[DNA_p]}_\mathrm{tot}\mathrm{[P \cdot DNA_p]}.
\end{multline}
Further rearrangment gives
\begin{multline}
\textcolor{black}{\alpha}\mathrm{[P \cdot DNA_p]}^2+\mathrm{[P \cdot DNA_p]}(-\mathrm{[P]}_\mathrm{tot}-\textcolor{black}{\alpha}\mathrm{[DNA_p]}_\mathrm{tot}-K_\mathrm{MD})+\mathrm{[P]}_\mathrm{tot}\mathrm{[DNA_p]}_\mathrm{tot} = 0.
\end{multline}
Solving the quadratic equation (and the constraint that the initiation complex can never exceed the concentration of each individual component) gives its analytical solution as
\begin{multline}
\mathrm{[P \cdot DNA_p]} = 
\frac{(\mathrm{[P]}_\mathrm{tot}+\alpha\mathrm{[DNA_p]}_\mathrm{tot}+K_\mathrm{MD})}{2 \alpha}
\\
-\frac{\sqrt{(\mathrm{[P]}_\mathrm{tot}+\alpha\mathrm{[DNA_p]}_\mathrm{tot}+K_\mathrm{MD})^2-4\alpha \mathrm{[P]}_\mathrm{tot}\mathrm{[DNA_p]}_\mathrm{tot}}}{2 \alpha}
\label{maincomplex}
\end{multline}
Finally, we recognize that the concentration of the DNA promoter is the same as that of the overall DNA sequence, 
\begin{equation}
    \mathrm{[DNA_p]_{tot}} = \mathrm{[DNA]_{tot}},
\end{equation}
which is the form used in the main text. 

\section{Analysis of TASEP Models}
\subsection{Model formulation}
The objective of this section is to describe our approach to developing a model that extends the initiation-elongation model to include the effects of polymerase-polymerase interactions during the elongation stage. We model the elongation stage as a totally asymmetric simple exclusion process (TASEP). The DNA sequence is approximated as a series of sequential segments that can be occupied by only one RNA polymerase at a time. Since the T7 RNA polymerase particles have a width that is not zero, the size of the segments is approximated as the width of the excluding polymerase molecules $L$. As such, the DNA sequence is divided into $M$ segments, where
\begin{equation}
    M = \frac{N_\mathrm{all}}{L}.
\end{equation}
In addition, we normalize the elongation rate constant to reflect this change,
\begin{equation}
    k_e' = \frac{k_{e\mathrm{,tot}}}{L}
\end{equation}The reaction sequence in the model is 
\begin{equation}
\mathrm{P + DNA_p} \mathrel{\mathop{\rightleftarrows}^{k_\mathrm{on}}_{k_\mathrm{off}} } 
\mathrm{P \cdot DNA_p}\xrightarrow
[-D]{k_{i}}{\mathrm{A_{1}}\xrightarrow
{k_\mathrm{e}'}\mathrm{A_{2}}\xrightarrow
{k_\mathrm{e}'}\cdots\xrightarrow
{k_\mathrm{e}'}\mathrm{A_{M}}\xrightarrow[-\mathrm{P}]
{k_\mathrm{e}'}}\,\mathrm{RNA}
\end{equation}
where $A_i$ represents the $i$th segment of the DNA sequence. TASEP modeling in the context of transcription has been studied before in the literature. We use the approach of \citeauthor{wang_minimal_2014} (\citeyear{wang_minimal_2014}), who develop and validate mean-field approximations to model the dynamics of TASEP systems relevant for transcription. The primary modifications of our approach is developing reasonable boundary conditions for the system, as \citeauthor{wang_minimal_2014} perform their analysis on a boundary-less system. The effective rate constant $k_e'$ used here includes the contribution of long pauses on the elongation rate of a single particle. The modeling approach presented below accounts for the effect of long pauses on transcription kinetics. We consider a model that neglects these long pauses in a later section.

The polymerase flow $R_p$ from the initiation complex to site $A_1$ is
\begin{equation}
    \frac{R_p}{\mathrm{[DNA_p]_{tot}}} = \frac{k_i\mathrm{[P \cdot DNA_p]}}{\mathrm{[DNA_p]_{tot}}} \frac{1-a_1}{1+a_1 k_e'\frac{f\tau^2}{1+f\tau}}
\end{equation}
where $a_k = \mathrm{[A_k]}/{\mathrm{[DNA_p]_{tot}}}$ and $R_p$ has units of M/h. The values $\tau$ and $1/f$ represent the timescales of pausing and unpausing, respectively. This equation can be conceptualized as a first-order rate law (the first term) with a set of correcting terms (the second term). The numerator of the right-hand fraction represents the probability that the movement of the particle is blocked by a previous particle in an active state. The denominator represents the probability that the particle is blocked by a previous particle in a paused state. Using a similar approach, the polymerase flow out of an inner site $A_k$ is
\begin{equation}
    \frac{R_p}{\mathrm{[DNA_p]_{tot}}} = k_e' a_k \frac{1-a_{k+1}}{1+a_{k+1} k_e'\frac{f\tau^2}{1+f\tau}}
\end{equation}
Finally, the polymerase flow from the last site ($A_M$) back into the bulk solution is 
\begin{equation}
    \frac{R_p}{\mathrm{[DNA_p]_{tot}}} = k_e' a_M.
\end{equation}
Other than these rates, all kinetic processes are the same as the initiation-elongation model. This set of rate equations could be modeled as a series of $M+3$ differential equations. However, considering that \citeauthor{wang_minimal_2014} only validate their mean-field approximations, it is more appropriate to evaluate this model only in the quasisteady limit. In theory, these equations could be solved numerically to give quasisteady transcription rates. However, the structure of these equations suggests that an (approximate) analytical solution is tractable. Below we derive an analytical version of the above model, which is useful for conceptual understanding of model behavior and is more amenable to parameter estimation calculations.

\subsection{Derivation of analytical TASEP model for transcription including long pauses}
For ease of derivation, define the dimensionless quantities
\begin{equation}
    p_\mathrm{tot} = \frac{\mathrm{[P]_{tot}}}{{\mathrm{[DNA_p]_{tot}}}},\  d = \frac{\mathrm{[DNA_p]}}{{\mathrm{[DNA_p]_{tot}}}},\  p = \frac{\mathrm{[P]}}{{\mathrm{[DNA_p]_{tot}}}},\  i = \frac{\mathrm{[P \cdot DNA_p]}}{{\mathrm{[DNA_p]_{tot}}}},\  a_k = \frac{\mathrm{[A_k]}}{{\mathrm{[DNA_p]_{tot}}}}.
\end{equation}
A mass balance around the DNA concentration gives 
\begin{equation}
    d = 1-i
\end{equation}
From previous work and numerical simulation of the model, it is understood that the interior segments, in the limit of long sequences, trend to a constant value. From direct simulation, we observe that the vast majority of sites have approximately the same polymerase density, which we call $a$ (Figure \ref{fig:constant}). We  assume at this stage that all sites have the same occupancy $a$. We will critically evaluate this assumption at a later stage. 

\begin{figure}[h] % use {figure} or non-float environment like {minipage} (the latter allows you to center a figure within an indented paragraph)
 \centering
 \includegraphics[width=0.8\textwidth]{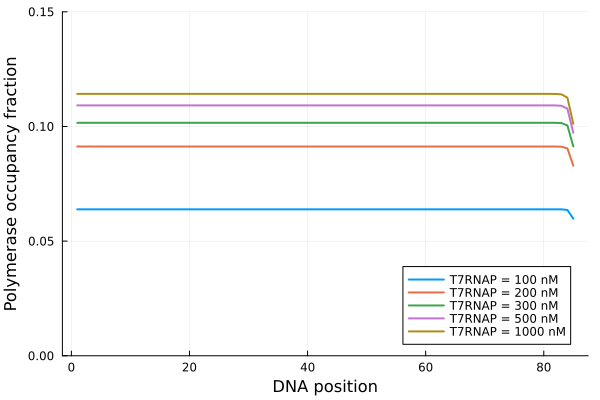}

 \vspace{-0.3cm}
 
\caption{Polymerase density as a function of segment number for different concentrations of T7RNAP. Values generated by numerical simulation for a model not considering long pauses. The initiation rate constant was set to 1500 1/h to reflect physically likely values. The segment density is roughly constant with the exception of a boundary layer at the tail of the sequence.} % caption to the figure
\label{fig:constant}
 \end{figure}

Using this approximation, the rate of polymerase flow per DNA molecule at some interior point is 
 \begin{equation}
 k_e' a\frac{1-a}{1+a k_e'\frac{f\tau^2}{1+f\tau}}
 \end{equation}
A flux balance is used to match the rate of initiation and elongation,
\begin{equation}
 k_ii\frac{1-a}{1+a k_e'\frac{f\tau^2}{1+f\tau}} =  k_e' a\frac{1-a}{1+a k_e'\frac{f\tau^2}{1+f\tau}},
\end{equation}
which results in a polymerase density of
\begin{equation}
 a = \frac{k_i}{k_e'}i.
\end{equation}

Performing a flux balance around the concentration of the initiation complex gives
\begin{equation}
k_\mathrm{on}pd{\mathrm{[DNA_p]_{tot}}}=k_\mathrm{off}i+k_ii\frac{1-a}{1+a k_e'\frac{f\tau^2}{1+f\tau}}.
\end{equation}
Inserting the above expressions for $d$ and $a$ and solving for $p$ gives that
\begin{equation}
    p=\frac{k_\mathrm{off}}{k_\mathrm{on}{\mathrm{[DNA_p]_{tot}}}}\frac{i}{1-i}+\frac{k_i}{k_\mathrm{on}{\mathrm{[DNA_p]_{tot}}}}\frac{i(1-\frac{k_i}{k_e'}i)}{(1-i)(1+i k_i\frac{f\tau^2}{1+f\tau})}.
\end{equation}

Defining the two constants
\begin{equation}
    \gamma = k_i\frac{f\tau^2}{1+f\tau},\qquad \theta = \frac{k_\mathrm{off}}{k_\mathrm{on}},
\end{equation}
simplifies the expression to
\begin{equation}
    p=\frac{\theta}{{\mathrm{[DNA_p]_{tot}}}}\frac{i}{1-i}+\frac{k_i}{k_\mathrm{on}{\mathrm{[DNA_p]_{tot}}}}\frac{i(1-\frac{k_i}{k_e'}i)}{(1-i)(1+\gamma i)}
    \label{imstartingpoint}
\end{equation}
A mass balance on the polymerase molecules is
\begin{equation}
    p_\mathrm{tot} = p + i + Ma = p + \alpha i,
    \label{eq:p-total}
\end{equation}
where 
\begin{equation}
    \alpha = 1+\frac{Mk_i}{k_e'} = 1+\frac{N_\mathrm{all}k_i}{k_e}.
\end{equation}
Substituting above equations into \eqref{eq:p-total} gives that
\begin{equation}
    p_\mathrm{tot} = \frac{\theta}{{\mathrm{[DNA_p]_{tot}}}}\frac{i}{1-i}+\frac{k_i}{k_\mathrm{on}{\mathrm{[DNA_p]_{tot}}}}\frac{i(1-\frac{k_i}{k_e'}i)}{(1-i)(1+\gamma i)} + \alpha i.
    \label{polynomialorigin}
\end{equation}
Rearrange to form a cubic equation in $i$:
\begin{equation}
    p_\mathrm{tot}(1-i)(1+\gamma i) = \frac{\theta}{{\mathrm{[DNA_p]_{tot}}}}i(1+\gamma i)+\frac{k_i}{k_\mathrm{on}{\mathrm{[DNA_p]_{tot}}}}i\big(1-\tfrac{k_i}{k_e'}i\big) + \alpha i(1-i)(1+\gamma i),
\end{equation}
\begin{multline}
\alpha \gamma i^3+ \Big(\alpha+\frac{k_i^2}{k_e'k_\mathrm{on}{\mathrm{[DNA_p]_{tot}}}} - \gamma\Big(p_\mathrm{tot} + \frac{\theta}{{\mathrm{[DNA_p]_\mathrm{tot}}}}+\alpha\Big)\Big)i^2\\-\Big(p_\mathrm{tot} + \alpha + \frac{\theta}{{\mathrm{[DNA_p]_\mathrm{tot}}}} + \frac{k_i}{k_\mathrm{on}{\mathrm{[DNA_p]_\mathrm{tot}}}} -\gamma p_\mathrm{tot}\Big)i +p_\mathrm{tot} = 0
\end{multline}
By defining the composite parameters
\begin{equation}
    K_{\mathrm{MD}} = \frac{k_i + k_\mathrm{off}}{k_\mathrm{on}} = \frac{k_i}{k_\mathrm{on}}+\theta,\qquad \beta = \frac{k_i^2}{k_e'k_\mathrm{on}} = \frac{Lk_i^2}{k_ek_\mathrm{on}},
\end{equation}
the expression is simplified to
\begin{equation}
\alpha \gamma i^3+\Big(\alpha+\frac{\beta}{{\mathrm{[DNA_p]_\mathrm{tot}}}} - \gamma\Big(p_\mathrm{tot} + \frac{\theta}{{\mathrm{[DNA_p]_\mathrm{tot}}}}+\alpha\Big)\Big)i^2-\Big(p_\mathrm{tot} + \alpha + \frac{K}{{\mathrm{[DNA_p]_{tot}}}} -\gamma p_\mathrm{tot}\Big)i+p_\mathrm{tot} = 0.
\label{poly}
\end{equation}
The value of $i$ in this cubic equation can be solved analytically. The reaction rate in nM/hr is equivalent to the initiation rate, whose analytical expressions are 
\begin{equation}
 \frac{R_p}{{\mathrm{[DNA_p]_{tot}}}} = k_ii\frac{1-a}{1+a k_e'\frac{f\tau^2}{1+f\tau}}  = k_ii\frac{1-\frac{\beta i}{K-\theta}}{1+\gamma i} 
 \label{poly2}
\end{equation}
Our final analytic expression for the transcription rate (equation \ref{poly} and rightmost expression of equation \ref{poly2}) is a function of the six parameters $k_i$, $\alpha$, $K$, $\beta$, $\theta$, and $\gamma$. %RDB: as stated, the expression is f a function of a (k_e') as well.
In the notation used in the main text of this paper, these rate laws can be written as
\begin{equation}
 R_p = k_i\mathrm{[P \cdot DNA_p]}\frac{({\mathrm{[DNA_p]_{tot}}}-\frac{\beta \mathrm{[P \cdot DNA_p]}}{K_{\mathrm{MD}}-\theta})}{{\mathrm{[DNA_p]_{tot}}}+\gamma \mathrm{[P \cdot DNA_p]}} 
\end{equation}
where 
\begin{multline}
\alpha \gamma \!\left(\frac{\mathrm{[P \cdot DNA_p]}}{{\mathrm{[DNA_p]_{tot}}}}\right)^{\!\!3} +\, \left(\alpha+\frac{\beta}{{\mathrm{[DNA_p]_{tot}}}} - \gamma\!\left(\frac{\mathrm{[P]_{tot}}}{{\mathrm{[DNA_p]_{tot}}}} + \frac{\theta}{{\mathrm{[DNA_p]_{tot}}}}+\alpha\right)\!\right)\left(\frac{\mathrm{[P \cdot DNA_p]}}{{\mathrm{[DNA_p]_{tot}}}}\right)^{\!\!2} \\ - \left(\frac{\mathrm{[P]_{tot}}}{{\mathrm{[DNA_p]_{tot}}}} + \alpha + \frac{K}{{\mathrm{[DNA_p]_{tot}}}} -\gamma \frac{\mathrm{[P]_{tot}}}{{\mathrm{[DNA_p]_{tot}}}}\right)\left(\frac{\mathrm{[P \cdot DNA_p]}}{{\mathrm{[DNA_p]_{tot}}}}\right)\!+\frac{\mathrm{[P]_{tot}}}{{\mathrm{[DNA_p]_{tot}}}} = 0
\end{multline}
where
\begin{equation}
    \gamma = k_i\frac{f\tau^2}{1+f\tau},\qquad \theta = \frac{k_\mathrm{off}}{k_\mathrm{on}}, \qquad\beta = \frac{Lk_i^2}{k_ek_\mathrm{on}}.
\end{equation}
\subsection{Evaluating approximations}
The earlier section assumed that all sites had the same occupancy $a$ based on an empirical observation from direct simulation. In this section, we evaluate this approximation and introduce a modification to account for errors generated by this approximation. We use a flux balance to match the polymerase flow off of the end of the sequence with the polymerase flow at a interior point,
\begin{equation}
    k_e' a_M = k_e' a\frac{1-a}{1+a k_e'\frac{f\tau^2}{1+f\tau}} = k_e' a\frac{ 1-a}{1+a \gamma \frac{K_{\mathrm{MD}}-\theta}{\beta}},
\end{equation}
and solve for $a$ to give
\begin{equation}
    a = \frac{\big(1-a_M\gamma \frac{K_{\mathrm{MD}}-\theta}{\beta}\big)-\sqrt{\big(1-a_M\gamma \frac{K_{\mathrm{MD}}-\theta}{\beta}\big)^{2}-4a_M}}{2}.
    \label{eq:a}
\end{equation}

\begin{figure}[h] % use {figure} or non-float environment like {minipage} (the latter allows you to center a figure within an indented paragraph)
 \centering
 \includegraphics[width=0.8\textwidth]{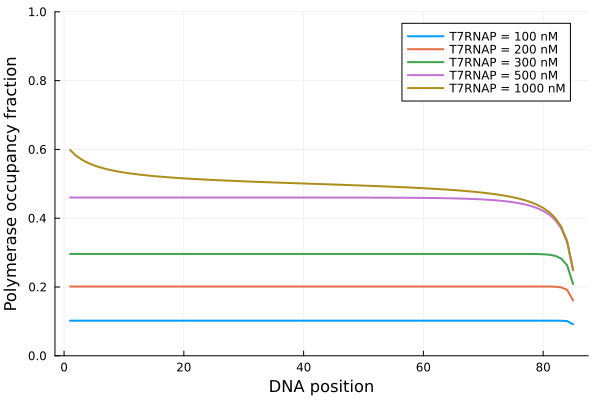}

 \vspace{-0.3cm}
 
\caption{Polymerase density as a function of segment number. Values generated by numerical simulation for a model not considering long pauses. The initiation rate constant was increased to 8500 1/h to demonstrate the operation of the model in a regime of very high fluxes. The final segment concentration does not exceed 0.25, and the concentration of interior points does not exceed 0.5. In cases where the concentration of the first segment exceeds 0.5, a boundary layer forms at the beginning of the sequence.} % caption to the figure
\label{fig:capping}
 \end{figure}

This expression gives some interesting conclusions: 
\begin{enumerate}
    \item $a_M$ can never be above 
    \begin{equation}
        a_{M,\max} = \left(\frac{\sqrt{\gamma \frac{K_{\mathrm{MD}}-\theta}{\beta}+1}-1}{\gamma \frac{K_{\mathrm{MD}}-\theta}{\beta}}\right)^{\!\!\!2}
    \end{equation}
    In the limit of $\gamma = 0$ (i.e., the TASEP model without long pauses), $a_{M,\max} =1/4$, which can be derived by applying L$^{\prime}$H\^{o}pital's rule.
    
    \item By insertion into \eqref{eq:a}, the upper bound on $a_M$ implies that $a$ has a maximum value of 
    \begin{equation}
        a_{\max} = \frac{\sqrt{\gamma \frac{K_{\mathrm{MD}}-\theta}{\beta}+1}-1}{\gamma \frac{K_{\mathrm{MD}}-\theta}{\beta}},
    \end{equation}
    which is 1/2 in the limit of $\gamma = 0$.

    With this in mind, we can reevaluate some of the steps of the earlier derivation. We assumed that the polymerase density of the first segment was equal to the polymerase density of the interior region. This assumption breaks when the polymerase density of the first segment rises above that of the maximum allowed by the interior. In this case, a boundary layer forms and the limiting step is the flux of polymerase molecules in the interior region (Figure \ref{fig:capping}). 

    In our code, we first test whether the polymerase density of the first segment exceeds the maximum value of the interior region. If this is not the case, we return values based on the main derivation presented. If this is the case, we return the interior flux for $a = a_{\max}$.

    We can assess the importance of this approximation by visually comparing results generated by direct simulation against our analytical approximation. Without adjusting for the effect of the boundary layer at the beginning of the sequence, the analytical approximation diverges from the results of numerical simulation in the limit of high RNA polymerase concentrations (Figure \ref{fig:compare}A). When the adjustment discussed in this section to account for boundary layer formation is made, this divergence is eliminated and the error of the analytical approximation is bounded below by roughly 1\%  (Figure \ref{fig:compare}).

\begin{figure}[h] % use {figure} or non-float environment like {minipage} (the latter allows you to center a figure within an indented paragraph)
 \centering
 \includegraphics[width=1.0\textwidth]{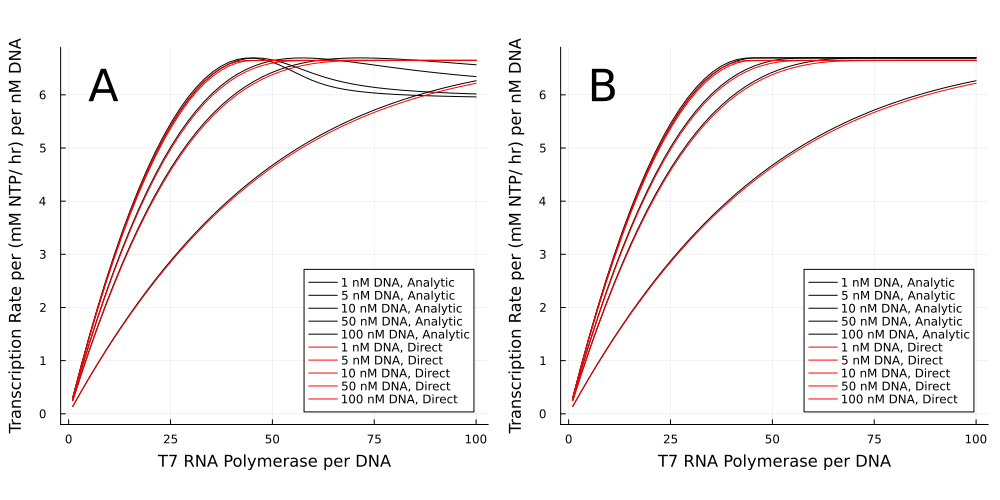}

 \vspace{-0.6cm}
 
\caption{Transcription rates as a function of DNA and polymerase concentrations generated by numerical simulation and analytical approximation. Analytical results are shown for a scheme that does not account for the formation of boundary layers (A) and a scheme that does (B). These results were generated with an elevated initiation rate constant of 8500 1/h to fully demonstrate differences between analytical and numerical results.} % caption to the figure
\label{fig:compare}
 \end{figure}
 
\end{enumerate}

\subsection{Alternate rate laws as special cases of TASEP with long pauses}
By considering special cases of the model presented above, simpler kinetic models (including the initiation-elongation and initiation-limited models) emerge as special cases. 
\subsubsection{TASEP model without long pauses}
First, consider the case in which pausing is relatively irrelevant to the system, which could be because 
\begin{enumerate}
    \item The fraction of paused particles is small.
    \item The timescale of pausing is small (recall that pauses with timescales faster than elongation are captured by the effective elongation constant).
    \item Particle density is low, and freely moving particles hardly ever interact with paused particles.
\end{enumerate}
The parameter $\gamma$ describes this case. Recall that 
\begin{equation}
    \gamma = k_i \tau \frac{f\tau}{1+f\tau} = \frac{\mathrm{timescale \ of \ pausing}}{\mathrm{timescale \ of \ initiation}}\mathrm{(fraction \ of \ particles \ in \ paused \ state)}
\end{equation}
So we would expect $\gamma$ to be nearly zero. This leads to the reduced expressions
\begin{equation}
\Big(\alpha+\frac{\beta}{{\mathrm{[DNA_p]_{tot}}}}\Big)i^2-\Big(p_\mathrm{tot} + \alpha + \frac{K}{{\mathrm{[DNA_p]_\mathrm{tot}}}}\Big)i+p_\mathrm{tot} = 0
\end{equation}
and 
\begin{equation}
  \frac{R_p}{{\mathrm{[DNA_p]_\mathrm{tot}}}} = k_ii\Big(1-\frac{\beta i}{K-\theta}\Big)
\end{equation}
\subsubsection{Initiation-elongation Model}
Secondly, consider the case in which particle interactions as a whole are irrelevant for the system. This would be because the total particle density is very low, which could be because
\begin{enumerate}
    \item The rate of initiation is very low.
    \item The rate of elongation is very high.
\end{enumerate}
This behavior is captured by the term
\begin{equation}
   \frac{\beta}{{\mathrm{[DNA_p]_{tot}}}} = \frac{k_i^2}{k_e'k_\mathrm{on}{\mathrm{[DNA_p]_{tot}}}} = \frac{\mathrm{Relative \ initiation \  rate}}{\mathrm{Relative \ elongation \ rate}} 
\end{equation}
As $\beta$ goes to zero, our model expressions simplify to 
\begin{equation}
\alpha i^2-\Big(p_\mathrm{tot} + \alpha + \frac{K}{{\mathrm{[DNA_p]_{tot}}}}\Big)i+p_\mathrm{tot} = 0
\end{equation}
and 
\begin{equation}
  \frac{R_p}{{\mathrm{[DNA_p]_{tot}}}} = k_ii,
\end{equation}
which is the initiation-elongation model discussed at length in this work.
\subsubsection{Initiation-limited Model}
Finally, consider the case in which the elongation phenomena is not necessary to capture the model. This could be because
\begin{enumerate}
    \item The rate of initiation is slow.
    \item The rate of elongation is very fast.
    \item The sequence is very short. 
\end{enumerate}
This behavior is captured by the parameter $\alpha$. Recall that
\begin{equation}
    \alpha = 1+\frac{N_\mathrm{all}k_i}{k_e} = \frac{\mathrm{Timescale \ of \ total \ transcription}}{\mathrm{Timescale \ of \ initiation\ }}
\end{equation}
In this initiation-limited regime, $\alpha$ approaches one, which further simplifies our expressions to
\begin{equation}
i^2-\Big(p_\mathrm{tot} + 1 + \frac{K}{{\mathrm{[DNA_p]_{tot}}}}\Big)i+p_\mathrm{tot} = 0
\end{equation}
and 
\begin{equation}
  \frac{R_p}{{\mathrm{[DNA_p]_{tot}}}} = k_ii.
\end{equation}

In summary, we have developed four competing models which are parameterized special cases of one another. This is a convenient starting point for model selection. 

\subsection{Parameter estimation of TASEP models}
The ability of the four models above to fit the Fluc data generated in this work, as well as synthetic data generated by the SP and LP models, was evaluated with Bayesian information criterion analysis. Table \ref{tab:BIC} compares the performance of these models. For this analysis, we considered the parameter $\theta$ to be fixed based on the values in Table \ref{tab:parameterfittingresults}, as the exact value of $\theta$ had minimal impact on the model predictions. 

\begin{table}[h!]
\begin{center}
\footnotesize
\label{table:BIC}
\caption{Comparison of BIC scores for the models. Models are compared on experimental data from the Fluc construct and synthetic data from the SP and LP models.}

\vspace{-0.2cm}

\begin{tabular}{||c c c c c||} 
 \hline
 Model & Parameters & BIC: Fluc Data & BIC: SP Syn. & BIC: LP Syn.\\ [0.5ex]
\hline\hline
Initiation-limited & $k_i, K_{MD}$ (with prior) & 500 & 380 & 270 \\   
\hline
Initiation-elongation & $k_i, K_{MD}$ (with prior), $\alpha$ & 62 & 40 & 43 \\   
\hline
Short-pause TASEP & $k_i, K_{MD}$ (with prior), $\alpha, \beta$ & 65 & 40 & 43 \\   
\hline
Long-pause TASEP & $k_i, K_{MD}$ (with prior), $\alpha, \beta, \gamma$ & 68 & 70 & 50 \\   
\hline
\end{tabular}
\label{tab:BIC}
\end{center}
\end{table}

As described in the main text, synthetic data generated by the LP and SP models was used to evaluate identifiability and distortion of the estimated initiation and elongation rate constants. Neither the LP or SP models could generate practically identifiable parameter confidence regions from fitting on synthetic data. This is best demonstrated in the companion code for this work available on Github. While the initiation-elongation model could reasonably fit the trends of LP and SP synthetic data, the resulting fitted kinetic parameters diverged from ground truth values. This divergence was most significant for the estimated initiation rate constant (Figure \ref{fig:paramest}).

\renewcommand{\arraystretch}{1.2}

\begin{table}[h!]

\begin{center}
\footnotesize
\label{table:parameterfittingresults}
\caption{Values of the microscopic parameters used in generating synthetic data for the TASEP models.}

\vspace{-0.2cm}

\begin{tabular}{||c c c c||} 
 \hline
 Parameter & Units & Source & Value\\ [0.5ex]
\hline\hline
$k_{i}$ & $\mathrm{h^{-1}}$ & This work &  1500 \\   
\hline
$k_{e,\mathrm{bp}}$ & $\mathrm{h^{-1}}$ & This work & $10^{5.5}$\\   
\hline
$k_\mathrm{on}$ & $\mathrm{h^{-1}}$ & \parencite{koh_correlating_2018} & $10^{2.30}$\\   
\hline
$k_\mathrm{off}$ & $\mathrm{h^{-1}}$ & \parencite{koh_correlating_2018} & $10^{3.74}$\\   
\hline
$\tau$ & $\mathrm{s}$ & \parencite{klumpp_growth-rate-dependent_2008} & 1\\   
\hline
$f$ & $\mathrm{h^{-1}}$ & \parencite{klumpp_growth-rate-dependent_2008} & $360$\\   
\hline
\end{tabular}
\label{tab:parameterfittingresults}
\end{center}
\end{table}
\renewcommand{\arraystretch}{1}

\begin{figure}[H] % use {figure} or non-float environment like {minipage} (the latter allows you to center a figure within an indented paragraph)
 \centering
 \includegraphics[width=0.7\textwidth]{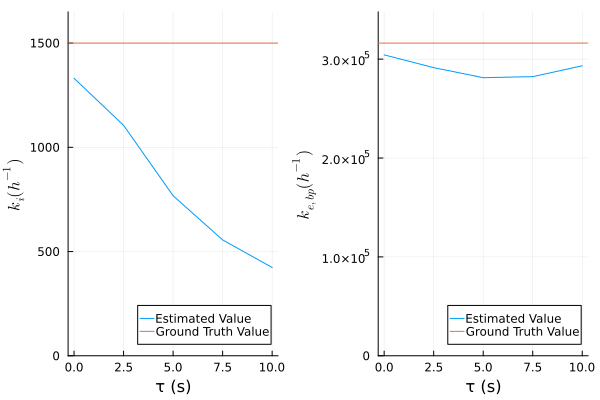}

 \vspace{-0.3cm}
 
\caption{Estimated rate constants of initiation-elongation model applied to data generated by long-pause TASEP model. Results are shown as a function of pausing time $\tau$. $\tau$ = 0 corresponds to short-pause TASEP model. While the estimated elongation rate constant does not diverge from ground truth value by more than 10\%, the estimated initiation rate constant is significantly influenced by pausing.} % caption to the figure
\label{fig:paramest}
 \end{figure}

\section{Salt Effects on IVT}
\subsection{First-principles model for polymerase-promoter binding in multi-salt systems}
Classical approaches to modeling the interactions of charged ligands and nucleic acids treat binding as the formation of $m$ ion pairs, which displace $n = m\psi$ ions \parencite{record_ion_1976}. This value $n$ is assumed to be constant in this work. We consider a thermodynamic ensemble at a constant temperature and chemical potential of salts in solution. With these assumptions, the binding reaction is written as
\begin{equation}
    \mathrm{P + DNA_p} \mathrel{\mathop{\rightleftarrows}^{K}} 
\mathrm{P \cdot DNA_p}
\end{equation}
At equilibrium, the chemical potentials of the three species can be related
\begin{equation}
    \left(\frac{\partial G}{\partial \xi}\right)_{\!T,V,{\mu}_s} = 0 = \mu_{\mathrm{P \cdot DNA_p}} - \mu_\mathrm{P} - \mu_{\mathrm{DNA_p}}
\end{equation}
For this derivation, the polymerase and initiation complex are assumed to behave as components in an ideal solution with a given reference chemical potential,
\begin{equation}
    \mu_\mathrm{P} = \mu^\circ_\mathrm{P}+k_BT \ln \frac{\mathrm{[P]}}{1 \mathrm{\,M}}
\end{equation}
\begin{equation}
    \mu_\mathrm{{P \cdot DNA_p}} = \mu^\circ_\mathrm{{P \cdot DNA_p}}+k_BT \ln \frac{\mathrm{[P \cdot DNA_p]}}{1 \mathrm{\,M}}
\end{equation}
While \citeauthor{record_ion_1976} \cite{record_ion_1976} incorporated salt effects by considering salt to be a reacting species, we describe salt effects by the form of the expression for the chemical potential of the unbound promoter. We assume that each cation-phosphate pairing decreases the chemical potential of $\mathrm{DNA_p}$,
\begin{equation}
    \mu_\mathrm{{D}} = \mu^\circ_\mathrm{{DNA_p}}+g_s + \ln \frac{\mathrm{[D]}}{1 \mathrm{\,M}}
\end{equation}
where $g_s$ is the free energy associated with cation-phosphate side binding,
\begin{equation}
    \frac{\mu^\circ_\mathrm{{P \cdot DNA_p}} - \mu^\circ_\mathrm{P} - \mu^\circ_\mathrm{D}}{k_BT} = \frac{\Delta \mu^\circ}{k_BT} = \ln \frac{(1\mathrm{\,M}) \mathrm{[P \cdot DNA_p]}}{\mathrm{[P][D]}} - \frac{g_s}{k_BT},
\end{equation}
\begin{equation}
    \ln K = \ln K_0 - \frac{g_s}{k_BT}.
\end{equation}
In the case of a single-salt system, the free energy of cation-phosphate binding is a combination of the cation chemical potential and an intrinsic binding energy $\epsilon$,
\begin{equation}
    \frac{g_s}{k_BT} = -\frac{n(\mu_s+\epsilon)}{k_BT}
    \label{singlesalt}
\end{equation}
where we model the salt species as a component of an ideal solution,
\begin{equation}
    \mu_s = \mu^\circ_s+k_BT \ln \frac{[S]}{1\mathrm{\,M}},
\end{equation}
\begin{equation}
    \frac{g_s}{k_BT} = \frac{n(\mu^\circ_s+\epsilon)}{k_BT}+n\ln \frac{[S]}{1\mathrm{\,M}},
\end{equation}
\begin{equation}
    \ln K = \ln K_0 - \frac{n(\mu^\circ_s+\epsilon)}{k_BT}+n\ln \frac{[S]}{1\mathrm{\,M}},
\end{equation}
leading to the same macroscopic predictions as classical counterion condensation theory,
\begin{equation}
    \ln K = \ln K_{0,\mathrm{obs}}+n\ln \frac{[S]}{1\mathrm{\,M}}.
\end{equation}
The derivation presented so far is mathematically identical to the approach of \citeauthor{record_ion_1976} \cite{record_ion_1976} in the case of ideal solution conditions (i.e., activity coefficents equal to one). Our reformulation, however, allows for a more natural extension to multiple-salt systems. In the case of a system with two different cations, the differential binding of these two cations must be considered. We model the DNA promoter as a Langmuir surface with a set number of binding sites $n$ in a system of constant temperature and chemical potential of each of the ion species. Each cation only occupies one binding site. The only accessible states are ones in which all sites are occupied. The partition function for this system is
\begin{equation}
    \Gamma = \left(\exp(\mu_{s,1}+\epsilon_1)+\exp(\mu_{s,2}+\epsilon_2)\right)^n
\end{equation}
where
\begin{equation}
    \frac{g_s}{k_BT} = \ln\Gamma = n \ln\!\left[\exp\!\left(\frac{\mu_{s,1}+\epsilon_1}{k_BT}\right)\!+\exp\!\left(\frac{\mu_{s,2}+\epsilon_2}{k_BT}\right)\right]
\end{equation}
which converges to \eqref{singlesalt} for the case of a single-cation system. In the case of two cations, 
\begin{equation}
    \frac{g_s}{k_BT} = \frac{n(\mu^\circ_{s,1}+\epsilon_1)}{k_BT}+n\ln\!\left(\frac{[S_1]+\omega[S_2]}{1\mathrm{\,M}}\right)
\end{equation}
where 
\begin{equation}
    \omega = \exp\left(\frac{\mu^\circ_{s,1}+\epsilon_1-\mu^\circ_{s,2}-\epsilon_2}{k_BT}\right),
\end{equation}
which recovers the semi-empirical observation of \citeauthor{lozinski_evaluation_2009} (\cite{lozinski_evaluation_2009}) that an ``effective salt concentration'' captures the binding trends in polymerase-DNA systems,
\begin{equation}
    \ln K = \ln K_{0,\mathrm{obs}}+n\ln\!\left(\frac{[\mathrm{S}_1]+\omega[\mathrm{S}_2]}{1\mathrm{\,M}}\right).
    \label{saltfinal}
\end{equation}
The above approach can be extended to anions (i.e., anion-polymerase binding) or to ions of different valence, to fully give physical explanation to the effective salt approach.

\subsection{Values of $\omega$ for estimating effective salt concentrations}
In this work, we quantify effective salt concentrations by summing over the ions in the system, in contrast to the approach of \citeauthor{lozinski_evaluation_2009} \cite{lozinski_evaluation_2009}, who sum over salt pairs. This change requires transformed values of $\omega$, which were calculated by assuming that the acetate ion has no effect on binding. 

\begin{table}[H]
\renewcommand{\arraystretch}{1.2}

\begin{center}
\caption{Values of $\omega$ in this work, adapted from \citeauthor{lozinski_evaluation_2009} 
 \cite{lozinski_evaluation_2009}.}

\vspace{-0.2cm}

\footnotesize
\label{table:parameterfittingresults}
\begin{tabular}{||c c||} 
 \hline
 Ion & $\omega$\\ [0.5ex]
\hline\hline
$\mathrm{Na^+}$ & 1\\   
\hline
$\mathrm{Mg^{2+}}$ & 4.71\\   
\hline
$\mathrm{HTris^+}$ & 1.07\\   
\hline
$\mathrm{Cl^-}$ & 0.72\\   
\hline
$\mathrm{OAc^-}$ & 0\\   
\hline
\end{tabular}
\label{tab:parameterfittingresults}
\end{center}
\renewcommand{\arraystretch}{1}
\end{table}

\section{Rate Laws for Double-stranded RNA Formation}
This section describes how the schematic mechanism in Figure \ref{fig:dsRNA} can be used to generate quantitative predictions of the proportion of dsRNA in IVT product. 
\subsection{Mechanism of 3' self-extension}
In the case of 3' self-extension, dsRNA is formed by a pathway of polymerase-undesired RNA promoter binding, forming an undesired initiation complex $(\mathrm{[P \cdot RNA_{up}]})$.
\begin{figure}[h] % use {figure} or non-float environment like {minipage} (the latter allows you to center a figure within an indented paragraph)
 \centering
 \includegraphics[width=0.95\textwidth]{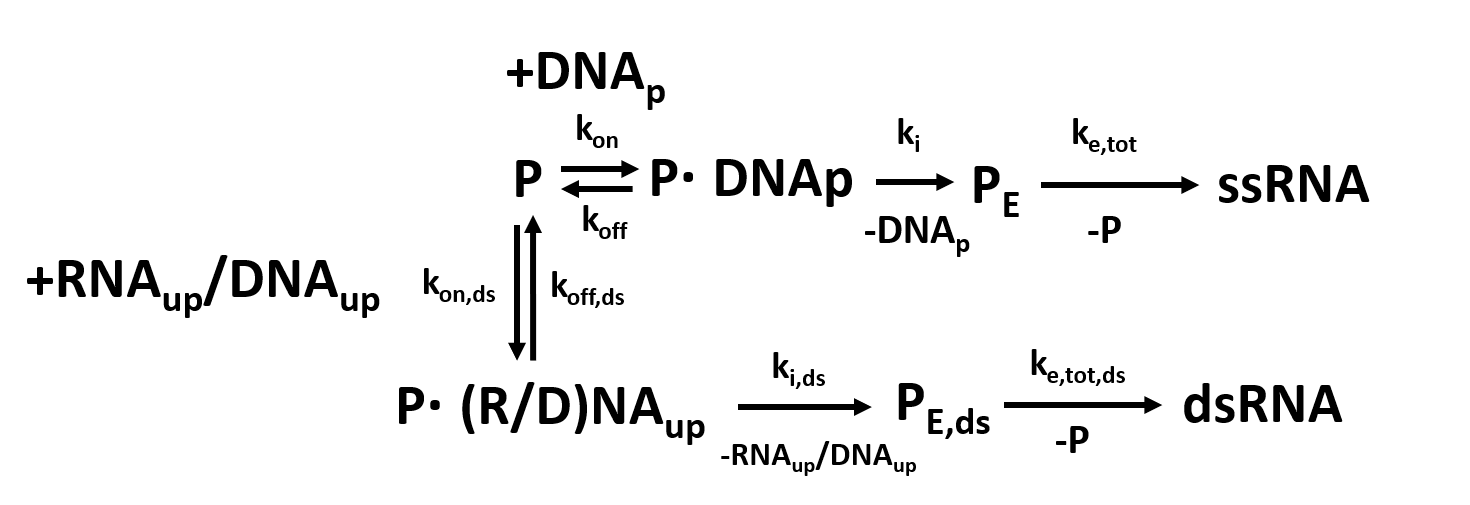}

\vspace{-0.4cm}
 
\caption{Mechanism for competition between double- and single-stranded RNA formation. For RNA-templated 3' extension, RNA polymerase binds to an undesired promoter on the RNA molecule. For DNA-templated antisense formation, the RNA polymerase instead binds to an undesired promoter on the antisense sequence of the template DNA.} % caption to the figure
\label{fig:dsRNA}
 \end{figure}

First note that 
\begin{equation}
    K_{\mathrm{MD}} = \frac{k_\mathrm{off}+k_\mathrm{on}}{k_i},
\end{equation}
\begin{equation}
    \mathrm{[P \cdot DNA_p]} = \frac{\mathrm{[P]}\mathrm{[DNA_p]}}{K_{\mathrm{MD}}}.
\end{equation}
Similarly, 
\begin{equation}
    \mathrm{[P \cdot RNA_{up}]} = \frac{\mathrm{[P]}\mathrm{[RNA_{up}]}}{K_{\mathrm{MD,R}}}
\end{equation}
where $\mathrm{[RNA_{up}]}$ is the free concentration of undesired promoter sites on RNA, and
\begin{equation}
    \mathrm{[P \cdot RNA_{up}]} = \frac{\mathrm{\mathrm{[RNA]_{tot}}}{K_{\mathrm{MD}}}}{\mathrm{[DNA_p]}K_{\mathrm{MD,R}}}\mathrm{[P \cdot DNA_p]}.
    \label{dsRNAcore}
\end{equation}
Also consider that 
\begin{equation}
    \mathrm{[P_{E,ds}]} = (1-\alpha_E)\mathrm{[P \cdot RNA_{up}]}
\end{equation}
where
\begin{equation}
    \alpha_\mathrm{E} = 1+\frac{k_{i,\mathrm{ds}}}{k_{e,\mathrm{tot,ds}}}
\end{equation}
The mass balance over the total concentration of polymerase is
\begin{equation}
\mathrm{[P]_{tot}} = \mathrm{[P]} + \mathrm{[P \cdot DNA_p]} + \mathrm{[P_{E}]} + \mathrm{[P \cdot RNA_{up}]} + \mathrm{[P_{E,ds}]}.
\end{equation}
Considering that the mass formation of dsRNA in a typical IVT reaction is around 1\%, we approximate that the dsRNA initiation and elongation complexes do not meaningfully contribute to the polymerase mass balance. Removing these species gives the same relation for the desired initiation complex concentration as \eqref{maincomplex}.

The rate of double-stranded RNA formation is 
\begin{equation}
k_{i,\mathrm{ds}}\mathrm{[P \cdot RNA_{up}]} = k_\mathrm{{ds}}\frac{\mathrm{[P \cdot DNA_p]}\mathrm{[RNA_{up}]}}{{\mathrm{[DNA_p]_{tot}}}-\mathrm{[P \cdot DNA_p]}}
\end{equation}
where
\begin{equation}
    k_\mathrm{{ds}} = \frac{k_{i,\mathrm{ds}}K_{\mathrm{MD}}}{K_{\mathrm{MD,R}}}.
\end{equation}
Based on previous sections, in the limit of high salt concentrations, the salt dependence of $k_{ds}$ is approximately
\begin{equation}
    k_{ds} \propto \frac{[\mathrm{salt}]^{n_f}}{[\mathrm{salt}]^{n_u}} = [\mathrm{salt}]^{n_f-n_u}
\end{equation}
where $n_f$ and $n_u$ represent the parameter $n$ of  \eqref{saltfinal} for the desired and undesired binding, respectively.

We are interested in the dsRNA concentration as a function of the extent of reaction. Some simplifying assumptions are
\begin{enumerate}
    \item All rate constants are constant as a function of the extent of reaction.

    \item The concentration of the undesired initiation complex $\mathrm{[P \cdot RNA_{up}]}$ is very small, relative both to the concentration of polymerase and the total concentration of RNA. This statement is equivalent to
    \begin{equation}
    \frac{\mathrm{\alpha_E}}{K_{\mathrm{MD,R}}}\ll 1,
    \end{equation}
   which allows for the statement
    \begin{equation}
        \mathrm{[RNA]_{tot}} \approx \mathrm{[RNA_{up}]}
    \end{equation}
    as well as the earlier assumption that the undesired promoter does not contribute meaningfully to the polymerase mass balance.
    This approximation is reasonable considering that dsRNA formation is typically a very small (0.1\%--1\%) fraction of the overall RNA concentration.

    \item Similarly to the above approximation, we assume that the final yield of single-stranded RNA is independent of dsRNA formation. This approximation is acceptable when the formation of dsRNA is very small in proportion to the formation of single-stranded RNA.
    
\end{enumerate}

Integrating the rate of dsRNA formation with respect to the extent of single-stranded RNA formation, the dsRNA concentration at some point of reaction conversion $f$ is
\begin{multline}
    \mathrm{[dsRNA]_f} = \int_{t=0}^{t=t_\mathrm{f}} k_\mathrm{{ds}}\frac{\mathrm{[P \cdot DNA_p]}\mathrm{[RNA_{up}]}}{{\mathrm{[DNA_p]_{tot}}}-\mathrm{[P \cdot DNA_p]}} dt\\ = \int_{\mathrm{[RNA]_{tot}}=0}^{\mathrm{[RNA]_{tot}}=\mathrm{[RNA]_{tot,f}}} \frac{k_\mathrm{{ds}}}{k_i}\frac{\mathrm{[RNA_{tot}]}}{{\mathrm{[DNA_p]_{tot}}}-\mathrm{[P \cdot DNA_p]}}d\mathrm{[RNA]_{tot}}\\ = \frac{k_\mathrm{{ds}}}{2k_i}\frac{\mathrm{\mathrm{[RNA]_{tot,f}}}^2}{\mathrm{[DNA_p]_{tot}}-\mathrm{[P \cdot DNA_p]}}
\end{multline}

This prediction also extends to a prediction for dsRNA fraction,
\begin{equation}
    \frac{\mathrm{[dsRNA]_f}}{\mathrm{\mathrm{[RNA]_{tot,f}}}} \propto     [\mathrm{salt}]^{n_f-n_u}\frac{\mathrm{\mathrm{[RNA]_{tot,f}}}}{{\mathrm{[DNA_p]_{tot}}}-\mathrm{[P \cdot DNA_p]}}.
\end{equation} 

\subsection{Mechanism of antisense transcription}

As an alternate mechanism for dsRNA formation, consider that RNA polymerase can bind to an undesired promoter on the antisense strand of the DNA $(\mathrm{[DNA_{up}]})$, forming an undesired antisense DNA initiation complex $\mathrm{(P \cdot DNA_{up})}$,
\begin{equation}
     \mathrm{[P \cdot DNA_{up}]} =  \frac{\mathrm{[P]}\mathrm{[DNA_{up}]}}{K_\mathrm{{MD,AS}}},
\end{equation} 
leading to
\begin{equation}
     \mathrm{[P \cdot DNA_{up}]} =  \frac{K_{\mathrm{MD}}\mathrm{[P \cdot DNA_p]}}{K_\mathrm{{MD,AS}}}\frac{\mathrm{[DNA_{up}]}}{\mathrm{[DNA_p]}}.
\end{equation}

This relation and \eqref{dsRNAcore} are  structurally similar. Using analogous assumptions, this allows us to write the rate of dsRNA formation as
\begin{equation}
k_{i,\mathrm{AS}}\mathrm{[P \cdot DNA_{up}]} = k_\mathrm{A/S}\frac{\mathrm{[P \cdot DNA_p]}\mathrm{[DNA_{up}]}}{{\mathrm{[DNA_p]_{tot}}}-\mathrm{[P \cdot DNA_p]}}
\end{equation}
where
\begin{equation}
    k_\mathrm{A/S} = \frac{k_{i,\mathrm{AS}}K_{\mathrm{MD}}}{K_\mathrm{{MD,AS}}}.
\end{equation}

Similarly to before, we assume that $\mathrm{[DNA_{up}]} \approx {\mathrm{[DNA_p]_{tot}}}$. Since the infinitesimal rate of dsRNA formation is constant with respect to the extent of reaction, an integral is unnecessary, and the expressions are
\begin{equation}
    \frac{\mathrm{[dsRNA]_f}}{\mathrm{\mathrm{[RNA]_{tot,f}}}} = \frac{k_\mathrm{A/S}}{k_i}\frac{{\mathrm{[DNA_p]_{tot}}}}{{\mathrm{[DNA_p]_{tot}}}-\mathrm{[P \cdot DNA_p]}}  \propto [\mathrm{salt}]^{n_f-n_u}\frac{{\mathrm{[DNA_p]_{tot}}}}{{\mathrm{[DNA_p]_{tot}}}-\mathrm{[P \cdot DNA_p]}},
\end{equation}
\begin{equation}
    \frac{\mathrm{[dsRNA]_f}}{\mathrm{\mathrm{[RNA]_{tot,f}}}} \propto     [\mathrm{salt}]^{n_f-n_u}\frac{{\mathrm{[DNA_p]_{tot}}}}{{\mathrm{[DNA_p]_{tot}}}-\mathrm{[P \cdot DNA_p]}}.
\end{equation}

\section{Effects of Conversion, Polymerase, and Salt on dsRNA Formation}
In addition to experiments measuring the dsRNA fraction as a function of IVT conversion performed on the Fluc construct described in the main text, similar measurements were performed on the COVID and EGFP constructs. No increasing trend was found in any of these data, however, a sequence dependence was apparent. No clear dependence of final dsRNA fraction on polymerase enzyme concentration was measured, subject to interpretation of an outlier point. %RDB: I am not sure that I agree with this statement. Other than the outlier, the models and the data show an increasing trend with polynomerase added. The difference is that the models predict a much higher dsRNA fraction. Is there a chemically justifable way to revise the algebraic relations in the model that would produce better agreement in panel B in Figure 6?

This is in contrast to the predictions of both RNA- and DNA-templated kinetic models. Similarly, no clear effect of NaCl addition was found on the final dsRNA fractions. This may be due to the narrow range of NaCl explored in these experiments. 
\begin{figure}[H] % use {figure} or non-float environment like {minipage} (the latter allows you to center a figure within an indented paragraph)
 \centering
 \includegraphics[width=1\textwidth]{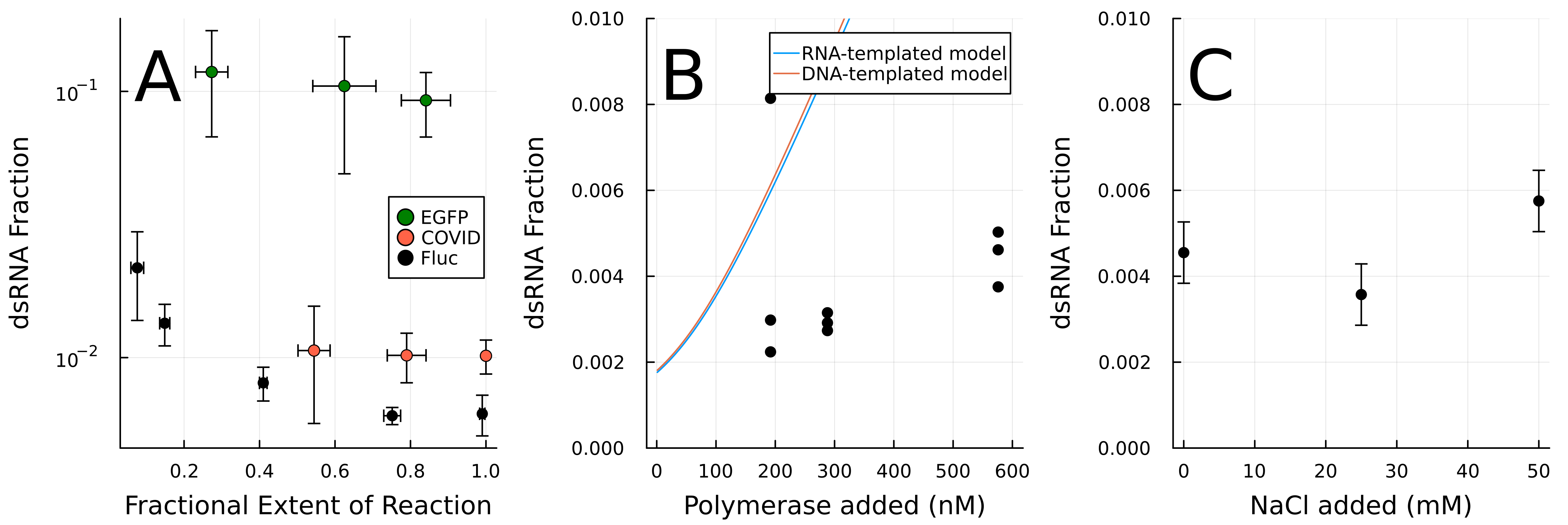}

\vspace{-0.2cm}
 
\caption{(A) Dependence of dsRNA fraction on reaction conversion. The DNA and RNA polymerase concentrations of all experiments were 9.2 and 192 nM, respectively. (B) Final dsRNA fraction of Fluc RNA as a function of polymerase concentration ([DNA] = 9.2 nM). (C) Final dsRNA fraction of Fluc RNA as a function of NaCl addition ([P] = 192 nM, [DNA] = 9.2 nM).} % caption to the figure
\label{fig:dsRNAresults}
 \end{figure}

\printbibliography
\end{document}